\documentclass[11pt,a4paper]{article}

\usepackage{keyval,amsfonts,slashed,bm}
\usepackage{graphicx,jheppub,subfigure,textcomp}
\usepackage[colorlinks=true,linktocpage=true,linkcolor=blue,citecolor=blue]{hyperref}

\def\sumint{\hbox{$\sum$}\!\!\!\!\!\!\!\int}
\def\sumintb{\sum\!\!\!\!\!\!\!\!\!\int\limits}
\def\sumintf{\sum\!\!\!\!\!\!\!\!\!\!\int\limits}
\def\sumintbb{\sum\!\!\!\!\!\!\!\!\!\!\int\limits}
\def\sumintbf{\sum\!\!\!\!\!\!\!\!\!\!\!\!\int\limits}
\def\sumintff{\sum\!\!\!\!\!\!\!\!\!\!\!\!\int\limits}
\def\sumintbbb{\sum\!\!\!\!\!\!\!\!\!\!\!\int\limits}
\def\sumintbbf{\sum\!\!\!\!\!\!\!\!\!\!\!\!\!\!\int\limits}
\def\sumintbff{\sum\!\!\!\!\!\!\!\!\!\!\!\!\!\!\int\limits}
\def\sumintfff{\sum\!\!\!\!\!\!\!\!\!\!\!\!\!\!\int\limits}
\def\bs{\!\!\!\!\!\!\!\!\!\!\!\!\!\!}
\def\bsa{\!\!\!\!\!\!\!\!\!}
\def\nn{\nonumber\\}

\def\lb{\left(}
\def\rb{\right)}
\def\hmu{\hat\mu}
\def\L{\ln\frac{\hat\Lambda}{2}}
\def\Lg{\ln\frac{\hat\Lambda_g}{2}}
\def\Za{\frac{\zeta'(-1)}{\zeta(-1)}}
\def\Zc{\frac{\zeta'(-3)}{\zeta(-3)}}
\def\be{\begin{eqnarray}}
\def\ee{\end{eqnarray}}
\def\del{\partial}
\def\[{\left[}
\def\]{\right]}

\begin{document}


\title{Three-loop HTLpt thermodynamics at finite temperature and chemical potential}


\author[a]{Najmul Haque,}
\author[a]{Aritra Bandyopadhyay,}
\author[b]{Jens O. Andersen,}
\author[a]{Munshi G. Mustafa,}
\author[c]{Michael Strickland,}
\author[d]{and Nan Su}
\affiliation[a]{Theory Division, Saha Institute of Nuclear Physics, 1/AF Bidhannagar, Kolkata-700107, India}
\affiliation[b]{Department of Physics, Norwegian University of Science and  Technology, N-7491 Trondheim, Norway}
\affiliation[c]{Department of Physics, Kent State University, Kent, Ohio 44242, United States}
\affiliation[d]{Faculty of Physics, University of Bielefeld, D-33615 Bielefeld, Germany}


\begin{abstract}{
We calculate the three-loop thermodynamic potential of QCD at finite temperature and chemical potential(s) using the hard-thermal-loop perturbation theory (HTLpt) reorganization of finite temperature and density QCD.  The resulting analytic thermodynamic potential allows us to compute the pressure, energy density, and entropy density of the quark-gluon plasma.  Using these we calculate the trace anomaly, speed of sound, and second-, fourth-, and sixth-order quark number susceptibilities.  For all observables considered we find good agreement between our three-loop HTLpt calculations and available lattice data for temperatures above approximately 300 MeV.
}
\end{abstract}


\preprint{BI-TP 2014/06}

\maketitle


\section{Introduction}
\label{intro}

Quantum chromodynamics (QCD) describes the propagation and interaction of quarks and gluons which are believed to be the fundamental constituents of all hadronic matter.  Based solely on the QCD Lagrangian it is possible to calculate the finite temperature and chemical potential partition function of QCD which results in the so-called equation of state (EoS).  The determination of the QCD EoS is extremely important to the phenomenology of the quark-gluon plasma (QGP).  At this time, the most reliable method to calculate the QCD thermodynamic functions at finite temperature and zero chemical potential is lattice gauge theory (see e.g.\cite{borsanyi1,borsanyi2,Borsanyi:2012uq,borsanyi3,borsanyi4,borsanyi5,sayantan,bnlb0,bnlb1,bnlb2,bnlb3,milc,hotqcd1,hotqcd2,peter_review}).  Importantly, lattice QCD can be used to probe the behavior of QCD matter near the transition temperature where QCD matter undergoes a phase transition from the hadronic phase to the deconfined QGP phase.  Near the phase transition, the running coupling is large and non-perturbative methods like lattice QCD must be used.  Finite temperature lattice QCD calculations are now quite sound; however, due to the sign problem, it is not straightforward to extend such calculations to finite baryon chemical potential.  In practice, it is possible to obtain information about the behavior of the thermodynamic functions at small baryon chemical potential by making a Taylor expansion of the partition function around $\mu_B=0$ and extrapolating the result.  This requires the calculation of various quark-number susceptibilities evaluated at zero chemical potential.  

Since extrapolations based on a finite number of Taylor coefficients can only be trusted within the radius of convergence of the expansion, it would be nice to have an alternative framework for calculating the finite temperature and chemical potential QCD thermodynamic potential and associated quantities.  This is important in light of the ongoing beam energy scan at the Relativistic Heavy Ion Collider (RHIC) and the forthcoming experiments at the Facility for Antiproton and Ion Research (FAIR).  As an alternative to lattice QCD calculations, one natural option is to compute the thermodynamic potential using perturbation theory.  In principle, this should work since, at sufficiently high temperature, the value of the strong coupling constant is small; however, one does not know a priori how large the temperature should be for this method to result in a good approximation to reality. The calculation of the thermodynamic potential using the weak-coupling expansion in the strong coupling constant, $g$, has a long history~\cite{shuryak,chin,kapusta79,toimela,arnoldzhai1,arnoldzhai2,zhaikastening,braatennieto1,braatennieto2} and the perturbative expansion of the pressure of QCD at both zero~\cite{kajantie} and non-zero chemical potential~\cite{vuorinen1,vuorinen2,ipp} are now known through order $g^6\ln g$. 

Unfortunately, it turns out that a strict expansion in the coupling constant converges only for temperatures many orders of magnitude higher than those relevant for heavy-ion collision experiments.  The source of the poor convergence comes from contributions from soft momenta, $p \sim g T$.  This suggests that one needs a way of reorganizing the perturbative series which treats the soft sector more carefully.  There are various ways of reorganizing the finite temperature/chemical potential perturbative series.  For scalar field theories one can use ``screened perturbation theory'' (SPT)~\cite{spt1,spt2,spt3,spt4,spt5} which was inspired in part by variational perturbation theory (VPT)~\cite{vpt1,vpt2,vpt3,vpt4,vpt5,vpt6}. For gauge theories, however, it is not possible to use a scalar gluon mass.  As a result, a gauge-invariant generalization of SPT called hard-thermal-loop perturbation theory (HTLpt) was developed.  HTLpt has been used to calculate thermodynamic functions at one loop HTLpt~\cite{andersen1,andersen2,andersen3,sylvain1,sylvain2}, at two loops~\cite{andersen4,andersen5,najmul2,najmul2qns}, and at three loops at zero chemical potential~\cite{3loopglue1,3loopglue2,3loopqed,3loopqcd1,3loopqcd2,3loopqcd3} as well as at finite chemical potential~\cite{najmul3}.  Application of some hard-thermal-loop motivated approaches can be found in~\cite{blaizotm3,blaizotm2,blaizotm1,blaizot1,blaizot2,blaizot3,najmul11,najmul12,najmul13,purnendu1,purnendu2,purnendu3}.  In addition to lattice QCD and HTLpt, there are also various model calculations on the market. For example, the Nambu-Jona-Lasinio (NJL)~\cite{njl1,njl2}, Polyakov-loop extended Nambu-Jona-Lasinio (PNJL)~\cite{pnjl1,pnjl2,pnjl3,pnjl4,pnjl5,pnjl6,pnjl7,pnjl8,pnjl9,pnjl10,pnjl11}, quasi-particle models~\cite{quasi1,quasi2,quasi3,quasi4,quasi5}, Polyakov-loop extended quark meson (PQM) model~\cite{pqm1,pqm2,pqm3} have been used to calculate various thermodynamic functions.  There have also been works which apply Pade and Borel-Pade methods to the perturbative QCD pressure \cite{pade1,pade2,pade3}.  Finally, we note that some results from holographic QCD for the quark number susceptibilities can be found in refs.~\cite{holographic1,holographic2}.

In this paper we calculate the thermodynamic potential at finite temperature and chemical potential to three-loop order in HTLpt.  The result for equal quark chemical potentials was first presented in ref.~\cite{najmul3}.  Herein, we present the details of this calculation and extend our results to the case that the quarks can possess flavor-dependent chemical potentials.  The resulting three-loop thermodynamic potential is renormalized using only known vacuum, mass, and coupling constant counterterms and the final result is completely analytic and gauge independent.  The resulting analytic thermodynamic potential is then used to obtain expressions for the pressure, energy density, entropy density, trace anomaly, speed of sound, and various quark number susceptibilities.  We find that there is good agreement between our NNLO HTLpt results and lattice data down to temperatures on the order of 300 MeV.

The paper is organized as follows.  In section~\ref{htlpt} we specify the HTLpt calculational framework and the necessary counterterms to renormalize HTLpt. In section~\ref{feyn_diag} we discuss the diagrams that contribute to the HTLpt thermodynamic potential through NNLO.  In section~\ref{sec:nnloomega} we present our final results for the NNLO thermodynamic potential.  In section~\ref{pres} we discuss the mass prescription for the in-medium masses $m_D$ and $m_q$.  We present our results for the thermodynamic functions and compare them with results from lattice gauge simulations in section~\ref{thermof}.  In section~\ref{sec:qns} we present our results for the \mbox{second-,} fourth-, and sixth-order baryon and quark number susceptibilities.  We also compare our results for these quantities with available lattice data.  In section~\ref{outlook} we summarize and conclude.  In appendix~\ref{expansion} the necessary diagrams are reduced to scalar sum-integrals and expanded in powers of $m_D/T$ and $m_q/T$.  We list the necessary non-trivial sum-integrals and integrals in appendices \ref{app:sum-integrals} and \ref{app:threedints}.  Finally, in appendix \ref{app:aleph} we list some properties of the $\aleph$ functions which appear repeatedly in finite density calculations.

\section{Hard-thermal-loop perturbation theory}
\label{htlpt}

The QCD Lagrangian density in Minkowski space can be written as
\be
{\cal L}_{\rm QCD}=-\frac{1}{2}{\rm Tr}[G_{\mu\nu}G^{\mu\nu}]+i\bar\psi\gamma^\mu D_{\mu}\psi+{\cal L}_{gh}+{\cal L}_{gf}
+\Delta{\cal L}_{\rm QCD} \, ,
\label{qcd_lag}
\ee
where the field strength is $G^{\mu\nu}=\partial^{\mu}A^{\nu}-\partial^{\nu}A^{\mu}-ig[A^{\mu},A^{\nu}]$ and the covariant derivative is $D^{\mu}=\partial^{\mu}-igA^{\mu}$.  The term $\Delta{\cal L}_{\rm QCD}$ contains the counterterms necessary to cancel ultraviolet divergences in perturbative calculations.  The ghost term ${\cal L}_{\rm gh}$ depends on the form of the gauge-fixing term ${\cal L}_{\rm gf}$. In this paper we work in general covariant gauge where ${\cal L}_{\rm gf} = - \xi^{-1} {\rm Tr}\left[\left(\partial_{\mu}A^{\mu}\right)^2\right]$ with $\xi$ being the gauge-fixing parameter.

Hard-thermal-loop perturbation theory is a reorganization of in-medium perturbation theory for QCD.  The HTLpt Lagrangian density can be written as
\be
 {\cal L}=\left.({\cal L}_{\rm QCD}+{\cal L}_{\rm HTL})\right|_{g\rightarrow\sqrt{\delta}g}+\Delta{\cal L}_{\rm HTL} \, , 
\label{total_lag} 
\ee
where the HTL improvement term is~\cite{lagrangian} 
\be
 {\cal L}_{\rm HTL}=(1-\delta)i m_q^2\bar\psi\gamma^\mu\left\langle\frac{y_\mu}{y\cdot\! D}\right\rangle_{\!\hat{\bf y}}\psi-\frac{1}{2}(1-\delta)
 m_D^2 {\rm Tr}\lb G_{\mu\alpha}\left\langle\frac{y^\alpha y_\beta}{(y\cdot\! D)^2}\right\rangle_{\!\hat{\bf y}} G^{\mu\beta}\rb \, ,
\label{htl_lag}
\ee
where $y^\mu = (1, {\bf\hat{y}})$ is a light-like four-vector with ${\bf\hat{y}}$ being a three-dimensional unit vector and the angular bracket indicates an average over the direction of ${\bf\hat{y}}$. The two parameters $m_D$ and $m_q$ can be identified with the Debye screening mass and the thermal quark mass, respectively, and account for screening effects.  HTLpt is defined by treating $\delta$ as a formal expansion parameter. By coupling the HTL improvement term (\ref{htl_lag}) to the QCD Lagrangian (\ref{qcd_lag}), HTLpt systematically shifts the perturbative expansion from being around an ideal gas of massless particles to being around a gas of massive quasiparticles which are the appropriate physical degrees of freedom at high temperature and/or chemical potential.

The HTLpt Lagrangian (\ref{total_lag}) reduces to the QCD Lagrangian (\ref{qcd_lag}) if we set $\delta=1$.  Physical observables are calculated in HTLpt by expanding in powers of $\delta$, truncating at some specified order, and then setting $\delta = 1$.  This defines a reorganization of the perturbative series in which the effects of $m_D^2$ and $m_q^2$ terms in (\ref{htl_lag}) are included to leading order but then systematically subtracted out at higher orders in perturbation theory by the $\delta m_D^2$ and $\delta m_q^2$ terms in (\ref{htl_lag}).  To obtain leading order (LO), next-to-leading order (NLO), and next-to-next-leading order (NNLO) results, one expands to orders $\delta^0$, $\delta^1$, $\delta^2$, respectively.  Note that HTLpt is gauge invariant order-by-order in the $\delta$ expansion and, consequently, the results obtained are independent of the gauge-fixing parameter $\xi$.

If the expansion in $\delta$ could be calculated to all orders, the final result would not depend on $m_D$ and $m_q$ when we set $\delta=1$. However, any truncation of the expansion in $\delta$ produces results that depend on $m_D$ and $m_q$. As a consequence, a prescription is required to determine $m_D$ and $m_q$ as a function of $T$, $\mu$ and $\alpha_s$. Several prescriptions had been discussed in~\cite{3loopqcd2} at zero chemical potential. The HTLpt expansion generates additional ultraviolet divergences. In QCD perturbation theory, renormalizability constrains the ultraviolet divergences to have a form that can be cancelled by the counterterm Lagrangian $\Delta {\cal L}_{\rm QCD}$ . We will demonstrate that the renormalization of HTLpt can be implemented by including a counterterm Lagrangian $\Delta{\cal L}_{\rm HTL}$ among the interaction terms in (\ref{htl_lag}). There is no all-order proof that the HTL perturbation expansion is renormalizable, so the general structure of the ultraviolet divergences is unknown. However, as shown previously in refs.~\cite{andersen4,andersen5,najmul2,3loopqcd2}, it is possible to renormalize the NNLO HTLpt thermodynamic potential using only a vacuum counterterm, a Debye mass counterterm, a fermion mass counterterm, and a coupling constant counterterm.  The necessary counterterms for renormalization of the NNLO thermodynamic potential are
\be
\Delta {\cal E}_0&=&\frac{d_A}{128\pi^2\epsilon}(1-\delta)^2m_D^4\ ,
\label{ctE}\\
\Delta m_D^2&=&\frac{11c_A-4s_F}{12\pi\epsilon}\alpha_s\delta(1-\delta)m_D^2\ ,
\label{ctmD}
\\
\Delta m_q^2&=&\frac{3}{8\pi\epsilon}\frac{d_A}{c_A}\alpha_s \delta(1-\delta)m_q^2\ ,
\label{ctmq}
\\
\delta\Delta\alpha_s&=&-\frac{11c_A-4s_F}{12\pi\epsilon}\alpha_s^2\delta^2\ ,
\label{ctalpha}
\ee
where, with the standard normalization, the QCD Casimir numbers are $c_A=N_c$, $d_A=N_c^2-1$, $s_F=N_f/2$, $d_F=N_cN_f$, and $s_{2F}=C_F s_f$ with $C_F = (N_c^2-1)/2N_c$.
Note that the coupling constant counterterm (\ref{ctalpha}) is consistent with one-loop running of $\alpha_s$.

In practice, in addition to the $\delta$ expansion, it is also necessary to make a Taylor expansion in the mass parameters scaled by the temperature, $m_D/T$ and $m_q/T$, in order to obtain analytically tractable sum-integrals. An added benefit of this procedure is that the final result obtained at NNLO is completely analytic. In order to truncate the series in $m_D/T$ and $m_q/T$ one treats these quantities as being ${\cal O}(g)$ at leading order, keeping all terms that naively contribute to the thermodynamic potential through ${\cal O}(g^5)$. In practice, such an truncated expansion works well \cite{sylvain2,spt5} and the radius of convergence of the scaled mass expansion seems to be quite large, giving us confidence in this approximate treatment of the necessary sum-integrals. 

In addition to calculations of the thermodynamic potential, hard-thermal-loop perturbation theory has been used to calculate various physical quantities which are relevant to the deconfined state of matter.  Quantities such as the dilepton production rate~\cite{dilepton1,dilepton2}, photon production rate~\cite{photon1}, single quark and quark anti-quark potentials~\cite{onetwo1,onetwo2,onetwo3,onetwo4,qqbar1,qqbar2,qqbar3,qqbar4,qqbar5}, fermion damping rate \cite{fermiondamp1,fermiondamp2,fermiondamp3}, photon damping rate~\cite{photondamping}, gluon damping rate~\cite{gluondamping1,gluondamping2}, jet energy loss~\cite{jetenergyloss1,jetenergyloss2,jetenergyloss3,jetenergyloss4,jetenergyloss5,jetenergyloss6,jetenergyloss7,jetenergyloss8,jetenergyloss9,jetenergyloss10,jetenergyloss11,jetenergyloss12}, plasma instabilities \cite{weibel1,weibel2,weibel3,weibel4,weibel5,weibel6,weibel7}, thermal axion production~\cite{axion}, and lepton asymmetry during leptogenesis \cite{lepton1,lepton2} have also been calculated using HTLpt.  We note, however, that most of the papers above have only worked at what we would call leading order in HTLpt.

\section{Contributions to the HTLpt thermodynamic potential through NNLO}
\label{feyn_diag}

The diagrams needed for the computation of the HTLpt thermodynamic potential through NNLO can be found
in figures 2 and 3 of ref.~\cite{3loopqcd2}.  In ref.~\cite{3loopqcd2} the authors computed the NNLO thermodynamic
potential at zero chemical potential.  Here we extend the NNLO calculation to finite chemical potential.\footnote{Some additional details concerning the LO and NLO finite chemical potential calculations can be found in refs.~\cite{sylvain1,sylvain2} and \cite{najmul2}.}  For this purpose, one needs to only consider diagrams which contain at least one quark propagator; however, for completeness we also list the purely gluonic contributions below.  In the results we will express thermodynamic quantities in terms of two dimensionless variables:  $\hat{m}_D = m_D/(2\pi T)$ and $\hat{\mu} = \mu/(2\pi T)$.

The complete NNLO HTLpt thermodynamic potential can be expressed in terms of these diagrams as
\begin{eqnarray}
 \Omega_{\rm NNLO}&=&d_A\left[{\cal F}_{1a}^g+{\cal F}_{1b}^g+{\cal F}_{2d}^g+{\cal F}_{3m}^g\right]+d_F\[{\cal F}_{1b}^f
                 +{\cal F}_{2d}^f+{\cal F}_{3i}^f\]\nonumber\\
&&\hspace{-.5cm}+d_Ac_A\Big[{\cal F}_{2a}^g+{\cal F}_{2b}^g+{\cal F}_{2c}^g+{\cal F}_{3h}^g+{\cal F}_{3i}^g+{\cal F}_{3j}^g
           +{\cal F}_{3k}^g+{\cal F}_{3l}^g\Big]   \nonumber\\
&&\hspace{-.5cm}+d_As_F\Big[{\cal F}_{2a}^f+{\cal F}_{2b}^f+{\cal F}_{3d}^f+{\cal F}_{3e}^f+{\cal F}_{3f}^f+{\cal F}_{3g}^f+
       {\cal F}_{3k}^f+{\cal F}_{3l}^f\Big]      \nonumber\\
&&\hspace{-.5cm}  +d_Ac_A^2\Big[{\cal F}_{3a}^g+{\cal F}_{3b}^g+{\cal F}_{3c}^g+{\cal F}_{3d}^g
                  +{\cal F}_{3e}^g+{\cal F}_{3f}^g+{\cal F}_{3g}^g\Big]+d_As_{2F}
         \Big[{\cal F}_{3a}^f+{\cal F}_{3b}^f\Big]   \nonumber\\
&&\hspace{-.5cm}+d_Ac_As_{F}
         \Big[-\frac{1}{2}{\cal F}_{3a}^f+{\cal F}_{3m}^f+{\cal F}_{3n}^f+{\cal F}_{3o}^f\Big]
         +d_As_F^2\Big[{\cal F}_{3c}^f+{\cal F}_{3j}^f\Big]   \nonumber\\
&&\hspace{-.5cm} + \Delta_0{\cal E}_0+\Delta_1{\cal E}_0+ \Delta_2{\cal E}_0+\Delta_1m_D^2\frac{\partial}
       {\partial m_D^2}\Omega_{\rm LO}+\Delta_1m_q^2\frac{\partial}{\partial m_q^2}\Omega_{\rm LO}
       \nonumber\\
&&\hspace{-.5cm} + \Delta_2m_D^2\frac{\partial}
       {\partial m_D^2}\Omega_{\rm LO} +\Delta_2m_q^2\frac{\partial}{\partial m_q^2}\Omega_{\rm LO}   +\Delta_1m_D^2\frac{\partial}{\partial m_D^2}
       \Omega_{\rm NLO}+\Delta_1m_q^2\frac{\partial}
       {\partial m_q^2}\Omega_{\rm NLO}
\nonumber \\
&& \hspace{-.5cm}
+\frac{1}{2}\left[\frac{\partial^2}{(\partial m_D^2)^2}\Omega_{\rm LO}
        \right]\left(\Delta_1 m_D^2\right)^2 +\frac{1}{2}\left[\frac{\partial^2}{(\partial m_q^2)^2}\Omega_{\rm LO}\right]\left(\Delta_1 m_q^2\right)^2
        \nonumber\\
&&\hspace{-.5cm}
+d_A\left[\frac{c_A{\cal F}_{2a+2b+2c}^g+s_F{\cal F}_{2a+2b}^f}{\alpha_s}\right]\Delta_1\alpha_s,
\end{eqnarray}
where the necessary counterterms at any order in $\delta$ can be calculated using eqs.~(\ref{ctE})-(\ref{ctalpha}).

The expressions for the one- and two-loop diagrams above can be found in refs.~\cite{andersen4,andersen5}.
The expressions for the three-loop bosonic diagrams ${\cal F}_{3a}^g$--${\cal F}_{3m}^g$ are presented in section 3 of ref.~\cite{3loopglue2}, and the three-loop diagrams with fermions ${\cal F}_{3a}^f$--${\cal F}_{3i}^f$ can be found in section 3 of ref.~\cite{3loopqed}. The three-loop diagrams specific to QCD, i.e., the non-Abelian diagrams involving quarks, are given by
\be\nonumber
{\cal F}_{\rm 3m}^f
&=& \frac{1}{6}
\sumint_{\{PQR\}}{\rm Tr}\left[
\Gamma^{\alpha}(R-P,R,P)S(P)\Gamma^{\beta}(P-Q,P,Q)S(Q)
\Gamma^{\gamma}(Q-R,Q,R)S(R)\right]
\\ &&\times
\Gamma^{\mu\nu\delta}(P-R,Q-P,R-Q)
\Delta^{\alpha\mu}(P-R)\Delta^{\beta\nu}(Q-P)\Delta^{\gamma\delta}(R-Q)\;,
\\
{\cal F}^f_{\rm 3n}&=&
{-}\sumint_{P}\bar{\Pi}_g^{\mu\nu}(P)\Delta^{\nu\alpha}(P)
\bar{\Pi}_f^{\alpha\beta}(P)\Delta^{\beta\mu}(P)\;,
\\
{\cal F}^f_{\rm 3o}&=&
-{1\over2}
g^2\sumint_{P\{Q\}}
{\rm Tr}\left[
\Gamma^{\alpha\beta}(P,-P,Q,Q)
S(Q)\right]\Delta^{\alpha\mu}(P)\Delta^{\beta\nu}(P)
\bar{\Pi}_g^{\mu\nu}(P)\;,
\ee
where
\be
\nonumber
\bar{\Pi}^{\mu\nu}_g(P)
&=&{1\over2}g^2\sumint_Q
\Gamma^{\mu\nu,\alpha\beta}(P,-P,Q,-Q)\Delta^{\alpha\beta}(Q)
\\ && \nonumber
+{1\over2}g^2\sumint_Q\Gamma^{\mu\alpha\beta}(P,Q,-P-Q)\Delta^{\alpha\beta}(Q)
\Gamma^{\nu\gamma\delta}(P,Q,-P-Q)\Delta^{\gamma\delta}(-P-Q)
\\&&
+g^2\sumint_Q{Q^{\mu}(P+Q)^{\nu}\over Q^2(P+Q)^2}\;,
\\ 
\bar{\Pi}^{\mu\nu}_f(P)
&=&
{-}g^2\sumint_{\{Q\}}{\rm Tr}\left[\Gamma^{\mu}(P,Q,Q-P)
S(Q)\Gamma^{\nu}(P,Q,Q-P)S(Q-P)\right]
\;.
\ee
Thus $\bar{\Pi}^{\mu\nu}(P)$ is the one-loop gluon self-energy with
HTL-resummed propagators and vertices as in ref.~\cite{3loopqcd2}:
\be
\bar{\Pi}^{\mu\nu}(P) = 
c_A\bar{\Pi}^{\mu\nu}_g(P)+s_F\bar{\Pi}^{\mu\nu}_f(P)\;.
\ee

\section{NNLO HTLpt thermodynamic potential}
\label{sec:nnloomega}

One can evaluate the sum-integrals necessary analytically by expanding in the ratios $m_D/T$ and $m_q/T$.  For details concerning this expansion and intermediate results, we refer the reader to appendix \ref{expansion}.  We consider first  the case that all quarks have the same chemical potential $\mu_f = \mu = \mu_B/3$ where $f$ is a flavor index and $\mu_f \in \{ \mu_u, \mu_d, \mu_s, \cdots, \mu_{N_f} \}$.  After presenting the steps needed for this case, we present the general result with separate chemical potentials for each quark flavor.

\subsection{NNLO result for equal chemical potentials}

When all quarks have the same chemical potential $\mu_f = \mu = \mu_B/3$ we can straightforwardly combine the results for the various sum-integrals.  In this case, the unrenormalized three-loop HTLpt thermodynamic potential is
\begin{eqnarray}
\frac{\Omega_{\rm 3 loop}}{\Omega_0}
&=& \frac{7}{4}\frac{d_F}{d_A}\lb1+\frac{120}{7}\hmu^2+\frac{240}{7}\hmu^4\rb
    +\frac{s_F\alpha_s}{\pi}\bigg[-\frac{5}{8}\left(1+12\hat\mu^2\right)\left(5+12\hat\mu^2\right)
    \nn
    &&+\frac{15}{2}\left(1+12\hat\mu^2\right)\hat m_D+\frac{15}{2}\bigg(\frac{1}{\epsilon}-1
   -\aleph(z)+4\ln{\frac{\hat\Lambda}{2}}-2\ln\hat m_D\Big)\hat m_D^3
      -90\hat m_q^2 \hat m_D\bigg]
\nn
&+& s_{2F}\left(\frac{\alpha_s}{\pi}\right)^2\bigg[\frac{15}{64}\bigg\{35-32\frac{\zeta'(-1)}
      {\zeta(-1)}+472 \hat\mu^2 + 384  \frac{\zeta'(-1)}{\zeta(-1)} \hat\mu ^2+1328  \hat\mu^4
      \nn      
      &&+ 64\Big(-36i\hat\mu\aleph(2,z)+6(1+8\hat\mu^2)\aleph(1,z)+3i\hat\mu(1+4\hat\mu^2)\aleph(0,z)\Big)\bigg\}
      \nn &&
      \hspace{5mm}
      -\frac{45}{2}\hat m_D\left(1+12\hat\mu^2\right)\bigg]
\nn
\end{eqnarray}
\begin{eqnarray}
&+& \left(\frac{s_F\alpha_s}{\pi}\right)^2\left[\frac{5}{4\hat m_D}\left(1+12\hat\mu^2\right)^2+30\left(1+12\hat\mu^2
      \right)\frac{\hat m_q^2}{\hat m_D}\right.\nn
      &&+\left.\frac{25}{24}\Bigg\{ \left(1 +\frac{72}{5}\hat\mu^2+\frac{144}{5}\hat\mu^4\right)\lb\frac{1}{\epsilon}
     +6\ln\frac{\hat\Lambda}{2}\rb
       + \frac{31}{10}+\frac{6}{5}\gamma_E - \frac{68}{25}\frac{\zeta'(-3)}{\zeta(-3)}\right.\nn
       &&+ \left.\frac{12}{5}(25+12\gamma_E)\hat\mu^2 + \frac{24}{5}(61+36\gamma_E)\hat\mu^4
        - \frac{8}{5}(1+12\hat\mu^2)\frac{\zeta'(-1)}{\zeta(-1)} \right.\nn
       &&-  \frac{144}{5}\Big[8\aleph(3,z)+3\aleph(3,2z) +12 i \hat\mu\,(\aleph(2,z)+\aleph(2,2z)) -12\hat\mu^2\aleph(1,2z)\nn
       &&-\left.i \hat\mu(1+12\hat\mu^2)\,\aleph(0,z)  
       - (3+20\hat\mu^2)\aleph(1,z)\Big]\Bigg\}\right.\nn       
       &&-\left.\frac{15}{2}\Bigg\{\left(\lb1+12\hat\mu^2\rb\lb\frac{1}{\epsilon}+4\L-2\ln\hat m_D\rb
       \right. \right. \nonumber \nn && \left.\left. 
       \hspace{5mm} + (1+12\hat\mu^2)\left(\frac{4}{3}
       -\aleph(z)\right)+24\aleph(1,z)\right)\Bigg\}\hat m_D\right]
\nn
&+& \left(\frac{c_A\alpha_s}{3\pi}\right)\left(\frac{s_F\alpha_s}{\pi}\right)\Bigg[\frac{15}{2\hat m_D}\lb1+12\hmu^2\rb
     -\frac{235}{32}\Bigg\{\bigg(1+\frac{792}{47}\hat\mu^2+\frac{1584}{47}\hat\mu^4\bigg)\lb\frac{1}{\epsilon}
     +6\ln\frac{\hat\Lambda}{2}\rb
     \nonumber\\
    &&+\frac{1809}{470}\left(1+\frac{8600}{603}\hat\mu^2+\frac{28720}{603}\hat\mu^4\right)
   -\frac{48 \gamma_E }{47}\lb1+12\hat\mu^2\rb-\frac{32}{47}\lb1+6\hat\mu^2\rb\frac{\zeta'(-1)}{\zeta(-1)}
    \nonumber\\
    &&-\frac{464}{235}\frac{\zeta'(-3)}{\zeta(-3)}-\frac{288 }{47}\lb1+12\hmu^2\rb\ln\hat m_D
   -\frac{288}{47}\Big[2i\hat\mu\aleph(0,z)
 \nonumber\\
    &&    
-\left(3+68\hat\mu^2\right)\aleph(1,z)
   +72i\hmu \aleph(2,z)+26\aleph(3,z)\Big]\Bigg\}
\nonumber\\
   &&+\frac{315}{8}\Bigg\{\lb1+\frac{132}{7}\hat\mu^2\rb\lb\frac{1}{\epsilon}+6\L-2\ln\hat m_D\rb+\frac{88}{21}+\frac{440}{7}\hmu^2
   +\frac{22}{7}\lb1+12\hmu^2\rb\gamma_E
   \nonumber\\
   &&-\frac{8}{7}\frac{\zeta'(-1)}{\zeta(-1)}+\frac{4}{7}\aleph(z)+\frac{264}{7}\aleph(1,z)\Bigg\}\hat m_D
+90\frac{\hat m_q^2}{\hat m_D}\Bigg] + \frac{\Omega_{\rm 3 loop,\rm{YM}}}{\Omega_0} \, ,
\label{threeloopunrenorm}
\end{eqnarray}
where $\Omega_0=-d_A\pi^2T^4/45$ and $\frac{\Omega_{\rm 3 loop,{\rm YM}}}{\Omega_0}$ is the pure Yang Mills unrenormalized three-loop thermodynamic potential~\cite{3loopqcd2}.  Above, $\aleph(z)=\Psi(z)+\Psi(z^*)$ with $z=1/2-i\hmu$ and $\Psi$ being the digamma function (see app.~\ref{app:aleph} for more details and useful properties of $\aleph(z)$).

The sum of all counterterms through order $\delta^2$ is
\begin{eqnarray}
\frac{\Delta\Omega}{\Omega_0} &=& \frac{\Delta\Omega_1+\Delta\Omega_2}{\Omega_0} \nn
&=& \frac{s_F\alpha_s}{\pi}\Bigg[- \frac{15}{2}
 \left(\frac{1}{\epsilon}+2\L - 
  2 \ln \hat m_D  \right) \hat m_D^3 \Bigg]
 \nonumber \\ 
&+& \left(\frac{c_A\alpha_s}{3\pi}\right)\left(\frac{s_F\alpha_s}{\pi}\right)\Bigg[
   \frac{235}{32}\Bigg\{\left(1+\frac{792}{47}\hat\mu^2+\frac{1584}{47}\hmu^4\right)\lb\frac{1}{\epsilon}
   +4\ln\frac{\hat\Lambda}{2}\rb+\frac{56}{47}\Za 
   \nonumber\\
   &+& \frac{149}{47}\lb1+\frac{2376}{149}\hmu^2+\frac{4752}{149}\hmu^4\rb  +\frac{1584}{47}\lb1
    +4\hmu^2\rb\aleph(1,z)+\frac{1056}{47}\Za\hmu^2\Bigg\}
   \nonumber\\
&-&\frac{315}{8}\Bigg\{\lb1+\frac{132}{7}\hmu^2\rb\lb\frac{1}{\epsilon}+4\ln\frac{\hat\Lambda}{2}
    -2\ln{\hat m_D}\rb
   -\frac{8}{7}\frac{\zeta'(-1)}{\zeta(-1)}+\frac{61}{21}+44\hmu^2
\nonumber\\
&+& \frac{264}{7}\aleph(1,z)\Bigg\}\hat m_D
\Bigg]+\left(\frac{s_F\alpha_s}{\pi}\right)^2 \Bigg[ - \frac{25}{24}
   \Bigg\{\lb1+\frac{72}{5}\hmu^2+\frac{144}{5}\hmu^4\rb\lb\frac{1}{\epsilon} + 4\L+ 3\rb 
\nonumber\\
&+&\frac{144}{5}\lb1+4\hmu^2\rb\aleph(1,z) + 
   \frac{8}{5}\lb1+12\hmu^2\rb\frac{\zeta'(-1)}{\zeta(-1)}\Bigg\}  
\nn
&+& \frac{15}{2}\Bigg\{\lb1+12\hmu^2\rb\lb\frac{1}{\epsilon}+4\L - 
   2 \ln \hat m_D + \frac{7}{3}\rb+24\aleph(1,z)\Bigg\}\hat m_D
\Bigg]
+ \frac{\Delta\Omega^{\rm YM}}{\Omega_0} \, ,
\nonumber \\
\label{deltaomega}
\end{eqnarray}
where $\Delta\Omega^{\rm YM}$ is the pure-glue three-loop HTLpt counterterm~\cite{3loopqcd2}
\be
\frac{\Delta\Omega^{\rm YM}}{\Omega_0} &=& \frac{45}{8\epsilon}\hat m_D^4+ \frac{495}{8} \lb\frac{c_A\alpha_s}{3 \pi}\rb\lb
\frac{1}{\epsilon}+2\Lg-2\ln\hat m_D\rb\hat m_D^3
\nn
&&+\lb\frac{c_A\alpha_s}{3\pi}\rb^2\Bigg[\frac{165}{16}\lb\frac{1}{\epsilon}+4\Lg+2+4\Za\rb
\nn
&&-\frac{1485}{8}\lb\frac{1}{\epsilon}+4\Lg-2\ln\hat m_D+\frac{4}{3}+2\Za\rb\hat m_D\Bigg] \, .
\ee
Adding the total three-loop HTLpt counterterm~(\ref{deltaomega}) to the unrenormalized three-loop HTLpt thermodynamic potential (\ref{threeloopunrenorm}) we obtain our final result for the NNLO HTLpt thermodynamic potential in the case that all quarks have the same chemical potential 
\begin{eqnarray}
\frac{\Omega_{\rm NNLO}}{\Omega_0}
&=& \frac{7}{4}\frac{d_F}{d_A}\lb1+\frac{120}{7}\hmu^2+\frac{240}{7}\hmu^4\rb
    -\frac{s_F\alpha_s}{\pi}\bigg[\frac{5}{8}\left(1+12\hat\mu^2\right)\left(5+12\hat\mu^2\right)
    \nn
    &&-\frac{15}{2}\left(1+12\hat\mu^2\right)\hat m_D-\frac{15}{2}\bigg(2\ln{\frac{\hat\Lambda}{2}-1
   -\aleph(z)}\Big)\hat m_D^3
      +90\hat m_q^2 \hat m_D\bigg]
\nn
&+& s_{2F}\left(\frac{\alpha_s}{\pi}\right)^2\bigg[\frac{15}{64}\bigg\{35-32\lb1-12\hmu^2\rb\frac{\zeta'(-1)}
      {\zeta(-1)}+472 \hat\mu^2+1328  \hat\mu^4\nn
      &&+ 64\Big(-36i\hat\mu\aleph(2,z)+6(1+8\hat\mu^2)\aleph(1,z)+3i\hat\mu(1+4\hat\mu^2)\aleph(0,z)\Big)\bigg\}
\nn
&-& \frac{45}{2}\hat m_D\left(1+12\hat\mu^2\right)\bigg] +\left(\frac{s_F\alpha_s}{\pi}\right)^2\left[\frac{5}{4\hat m_D}\left(1+12\hat\mu^2\right)^2+30\left(1+12\hat\mu^2
        \right)\frac{\hat m_q^2}{\hat m_D}\right.\nn
        &+&\left.\frac{25}{12}\Bigg\{ \left(1 +\frac{72}{5}\hat\mu^2+\frac{144}{5}\hat\mu^4\right)\ln\frac{\hat\Lambda}{2}
       + \frac{1}{20}\lb1+168\hmu^2+2064\hmu^4\rb+\frac{3}{5}\lb1+12\hmu^2\rb^2\gamma_E \right.\nn
       && \left.
        - \frac{8}{5}(1+12\hat\mu^2)\frac{\zeta'(-1)}{\zeta(-1)} - \frac{34}{25}\frac{\zeta'(-3)}{\zeta(-3)}\right.
       -  \frac{72}{5}\Big[8\aleph(3,z)+3\aleph(3,2z)-12\hat\mu^2\aleph(1,2z)\nonumber
%
\\  
&& +12 i \hat\mu\,(\aleph(2,z)+\aleph(2,2z)) 
       -\left.i \hat\mu(1+12\hat\mu^2)\,\aleph(0,z)  
       - 2(1+8\hat\mu^2)\aleph(1,z)\Big]\Bigg\}\right.\nn
       &&-\left.\frac{15}{2}\Bigg\{\lb1+12\hat\mu^2\rb\lb2\L-1-\aleph(z)\rb\Bigg\}\hat m_D\right]
\nn
&+& \left(\frac{c_A\alpha_s}{3\pi}\right)\left(\frac{s_F\alpha_s}{\pi}\right)\Bigg[\frac{15}{2\hat m_D}\lb1+12\hmu^2\rb
     -\frac{235}{16}\Bigg\{\bigg(1+\frac{792}{47}\hat\mu^2+\frac{1584}{47}\hat\mu^4\bigg)\ln\frac{\hat\Lambda}{2}
     \nonumber\\
    &&-\frac{144}{47}\lb1+12\hmu^2\rb\ln\hat m_D+\frac{319}{940}\left(1+\frac{2040}{319}\hat\mu^2+\frac{38640}{319}\hat\mu^4\right)
   -\frac{24 \gamma_E }{47}\lb1+12\mu^2\rb
\nonumber\\
    &&
   -\frac{44}{47}\lb1+\frac{156}{11}\hmu^2\rb\frac{\zeta'(-1)}{\zeta(-1)}
    -\frac{268}{235}\frac{\zeta'(-3)}{\zeta(-3)}
   -\frac{72}{47}\Big[4i\hat\mu\aleph(0,z)
    \nonumber\\
    &&+\left(5-92\hat\mu^2\right)\aleph(1,z)+144i\hmu\aleph(2,z)
   +52\aleph(3,z)\Big]\Bigg\}+90\frac{\hat m_q^2}{\hat m_D}
\nonumber\\
   &&+\frac{315}{4}\Bigg\{\lb1+\frac{132}{7}\hmu^2\rb\L+\frac{11}{7}\lb1+12\hmu^2\rb\gamma_E+\frac{9}{14}\lb1+\frac{132}{9}\hmu^2\rb
\nn  
 &&+\frac{2}{7}\aleph(z)\Bigg\}\hat m_D 
\Bigg]
+ \frac{\Omega_{\rm NNLO}^{\rm YM}}{\Omega_0} \, .
\label{finalomega1}
\end{eqnarray}
where $\Omega_{\rm NNLO}^{\rm YM}$ is the NNLO pure-glue thermodynamic potential~\cite{3loopglue2}~\footnote{Note that chemical potential dependence also appears in pure-glue diagrams from the
internal quark loop in effective gluon propagators and effective vertices. This chemical potential dependence enters through the chemical potential dependence of the Debye mass.}
{\small
\be
\frac{\Omega_{\rm NNLO}^{\rm YM}}{\Omega_0} &=& 1-\frac{15}{4}\hat m_D^3+\frac{c_A\alpha_s}{3\pi}\Bigg[-\frac{15}{4}
+\frac{45}{2}\hat m_D-\frac{135}{2}\hat m_D^2-\frac{495}{4}\lb\Lg+\frac{5}{22}+\gamma_E\rb  \hat m_D^3 \Bigg]
\nn
&+&\lb\frac{c_A\alpha_s}{3\pi}\rb^2\Bigg[\frac{45}{4\hat m_D}-\frac{165}{8}\lb\Lg-\frac{72}{11}\ln\hat m_D-\frac{84}{55}-\frac{6}{11}
\gamma_E-\frac{74}{11}\Za+\frac{19}{11}\Zc\rb
\nn
&+&\frac{1485}{4}\lb\Lg-\frac{79}{44}+\gamma_E+\ln2-\frac{\pi^2}{11}\rb\hat m_D\Bigg] \, ,
\label{ymomega1}
\ee
}
The result contained in eq.~(\ref{finalomega1}) was first presented in ref.~\cite{najmul3}.  Note that the full thermodynamic potential (\ref{finalomega1}) reduces to thermodynamic potential of ref.~\cite{3loopqcd2} in the limit $\mu\rightarrow 0$.  In addition, the above thermodynamic potential produces the correct ${\cal O}(g^5)$ perturbative result when expanded in a strict power series in $g$~\cite{vuorinen1,vuorinen2}.\footnote{There is a mismatch in one term proportional to $s_{2F}\alpha_s^2$ compared to result published in refs.~\cite{vuorinen1,vuorinen2}. We found that the second term proportional to $s_{2F}\alpha_s^2$ is $32(1-12\hmu^2)\zeta'(-1)/\zeta(-1)$, whereas in refs.~\cite{vuorinen1,vuorinen2} it was listed as $32(1-4\hmu^2)\zeta'(-1)/\zeta(-1)$.  The author of refs.~\cite{vuorinen1,vuorinen2} has agreed that this was a typo in his article.}

\subsection{NNLO result -- General case}

It is relatively straightforward to generalize the previously obtained result (\ref{finalomega1}) to the case that each quark has a separate chemical potential $\mu_f$.  The final result is
\begin{eqnarray}
\frac{\Omega_{\rm NNLO}}{\Omega_0}
&=& \frac{7}{4}\frac{d_F}{d_A}\frac{1}{N_f}\sum\limits_f\lb1+\frac{120}{7}\hmu_f^2+\frac{240}{7}\hmu_f^4\rb
    -\frac{s_F\alpha_s}{\pi}\frac{1}{N_f}\sum\limits_f\bigg[\frac{5}{8}\left(1+12\hat\mu_f^2\right)\left(5+12\hat\mu_f^2\right)
    \nn
    &&-\frac{15}{2}\left(1+12\hat\mu_f^2\right)\hat m_D-\frac{15}{2}\bigg(2\ln{\frac{\hat\Lambda}{2}-1
   -\aleph(z_f)}\Big)\hat m_D^3
      +90\hat m_q^2 \hat m_D\bigg]
\nn
&&+ \frac{s_{2F}}{N_f}\left(\frac{\alpha_s}{\pi}\right)^2\sum\limits_f\bigg[\frac{15}{64}\bigg\{35-32\lb1-12\hmu_f^2\rb\frac{\zeta'(-1)}
      {\zeta(-1)}+472 \hat\mu_f^2+1328  \hat\mu_f^4\nn
      &&+ 64\Big(-36i\hat\mu_f\aleph(2,z_f)+6(1+8\hat\mu_f^2)\aleph(1,z_f)+3i\hat\mu_f(1+4\hat\mu_f^2)\aleph(0,z_f)\Big)\bigg\}\nn
      &&- \frac{45}{2}\hat m_D\left(1+12\hat\mu_f^2\right)\bigg] \nn
&&+ \left(\frac{s_F\alpha_s}{\pi}\right)^2
      \frac{1}{N_f}\sum\limits_{f}\frac{5}{16}\Bigg[96\left(1+12\hat\mu_f^2\right)\frac{\hat m_q^2}{\hat m_D}
     +\frac{4}{3}\lb1+12\hmu_f^2\rb\lb5+12\hat\mu_f^2\rb
      \ln\frac{\hat{\Lambda}}{2}\nn
    && +\frac{1}{3}+4\gamma_E+8(7+12\gamma_E)\hat\mu_f^2+112\mu_f^4-\frac{64}{15}\frac{\zeta^{\prime}(-3)}{\zeta(-3)}-
   \frac{32}{3}(1+12\hat\mu_f^2)\frac{\zeta^{\prime}(-1)}{\zeta(-1)}\nn
   &&-    96\Big\{8\aleph(3,z_f)+12i\hat\mu_f\aleph(2,z_f)-2(1+2\hat\mu_f^2)\aleph(1,z_f)-i\hat\mu_f\aleph(0,z_f)\Big\}\Bigg] \nn \nonumber
\end{eqnarray}
\begin{eqnarray}
&&+ \left(\frac{s_F\alpha_s}{\pi}\right)^2
      \frac{1}{N_f^2}\sum\limits_{f,g}\left[\frac{5}{4\hat m_D}\left(1+12\hat\mu_f^2\right)\left(1+12\hat\mu_g^2\right)
     +90\Bigg\{ 2\left(1 +\gamma_E\right)\hat\mu_f^2\hat\mu_g^2
      \right.\nn
        &&-\Big\{\aleph(3,z_f+z_g)+\aleph(3,z_f+z_g^*)+ 4i\hat\mu_f\left[\aleph(2,z_f+z_g)+\aleph(2,z_f+z_g^*)\right]-4\hat\mu_g^2\aleph(1,z_f)\nn
       &&
       -(\hat\mu_f+\hat\mu_g)^2\aleph(1,z_f+z_g)- (\hat\mu_f-\hat\mu_g)^2\aleph(1,z_f+z_g^*)-4i\hat\mu_f\hat\mu_g^2\aleph(0,z_f)\Big\}\Bigg\}\nn
       &&-\left.\frac{15}{2}\lb1+12\hat\mu_f^2\rb\lb2\L-1-\aleph(z_g)\rb  \hat m_D\right]
\nn
&&+ \left(\frac{c_A\alpha_s}{3\pi}\right)\left(\frac{s_F\alpha_s}{\pi N_f}\right)\sum\limits_f\Bigg[\frac{15}{2\hat m_D}\lb1+12\hmu_f^2\rb
     -\frac{235}{16}\Bigg\{\bigg(1+\frac{792}{47}\hat\mu_f^2+\frac{1584}{47}\hat\mu_f^4\bigg)\ln\frac{\hat\Lambda}{2}
     \nonumber\\
    &&-\frac{144}{47}\lb1+12\hmu_f^2\rb\ln\hat m_D+\frac{319}{940}\left(1+\frac{2040}{319}\hat\mu_f^2+\frac{38640}{319}\hat\mu_f^4\right)
   -\frac{24 \gamma_E }{47}\lb1+12\hat\mu_f^2\rb
\nonumber\\
    &&
   -\frac{44}{47}\lb1+\frac{156}{11}\hmu_f^2\rb\frac{\zeta'(-1)}{\zeta(-1)}
    -\frac{268}{235}\frac{\zeta'(-3)}{\zeta(-3)}
   -\frac{72}{47}\Big[4i\hat\mu_f\aleph(0,z_f)+\left(5-92\hat\mu_f^2\right)\aleph(1,z_f)
    \nonumber\\
    &&+144i\hmu_f\aleph(2,z_f)
   +52\aleph(3,z_f)\Big]\Bigg\}+90\frac{\hat m_q^2}{\hat m_D}+\frac{315}{4}\Bigg\{\lb1+\frac{132}{7}\hmu_f^2\rb\L
\nonumber\\
   &&+\frac{11}{7}\lb1+12\hmu_f^2\rb\gamma_E+\frac{9}{14}\lb1+\frac{132}{9}\hmu_f^2\rb
+\frac{2}{7}\aleph(z_f)\Bigg\}\hat m_D 
\Bigg]
+ \frac{\Omega_{\rm NNLO}^{\rm YM}}{\Omega_0} \, ,
\label{finalomega}
\end{eqnarray}
where the sums over $f$ and $g$ include all quark flavors, $z_f = 1/2 - i \hat{\mu}_f$, and $\Omega_{\rm NNLO}^{\rm YM}$ is the pure-glue contribution as before. The result contained in eq.~(\ref{finalomega}) is new compared to what was reported in ref.~\cite{najmul3} since it includes separate chemical potentials for all quark flavors.

\section{Mass prescription}
\label{pres}

As discussed in ref.~\cite{3loopqcd2}, the two-loop perturbative electric gluon mass, first introduced by Braaten and Nieto in ~\cite{braatennieto1,braatennieto2} is the most suitable for three-loop HTLpt calculations. We use the Braaten-Nieto (BN) mass prescription for $m_D$ in the remainder of the paper.  Originally, the two-loop perturbative mass was calculated in refs.~\cite{braatennieto1,braatennieto2} for zero chemical potential, however, Vuorinen has generalized it to finite chemical potential.  The resulting expression for $m_D^2$ is \cite{vuorinen1,vuorinen2}
\begin{eqnarray}
\hat m_D^2&=&\frac{\alpha_s}{3\pi} \Biggl\{c_A
+\frac{c_A^2\alpha_s}{12\pi}\lb5+22\gamma_E+22\Lg\rb +
\frac{1}{N_f} \sum\limits_{f}
\Biggl[ s_F\lb1+12\hmu_f^2\rb
\nonumber\\
  &&+\frac{c_As_F\alpha_s}{12\pi}\lb\lb9+132\hmu_f^2\rb+22\lb1+12\hmu_f^2\rb\gamma_E+2\lb7+132\hmu_f^2\rb\L+4\aleph(z_f)\rb
\nonumber\\  
&&+\frac{s_F^2\alpha_s}{3\pi}\lb1+12\hmu_f^2\rb\lb1-2\L+\aleph(z_f)\rb
 -\frac{3}{2}\frac{s_{2F}\alpha_s}{\pi}\lb1+12\hmu_f^2\rb \Biggr] \Biggr\} \, .
\end{eqnarray}
The effect of the in-medium quark mass parameter $m_q$ in thermodynamic functions is small and following
ref.~\cite{3loopqcd2} we take $m_q=0$ which is the three-loop variational solution.  The maximal effect on the susceptibilities comparing the perturbative quark mass, \mbox{$\hat{m}_q^2 = \alpha_s(1+4\hat\mu^2)/6\pi$}, with the variational solution, $m_q=0$, is approximately 0.2\% at $T=200$ MeV.  At higher temperatures, the effect is much smaller, e.g. 0.02\% at $T=1$ GeV.

\section{Thermodynamic functions} 
\label{thermof}

In this section we present our final results for the NNLO HTLpt pressure, energy density, entropy density,
trace anomaly, and speed of sound.

\subsection{Running coupling} 

Below we will generally use the self-consistent one-loop running coupling implied by eq.~(\ref{ctalpha}), however, in some places we will try to gauge the sensitivity of the result to the order of the running coupling by comparing the impact of using one- or three-loop running.  The three-loop running coupling can be expressed approximately as~\cite{alphas1,alphas2}~\footnote{We have checked that for the scale range of interest, this is a very good approximation to the exactly integrated QCD three-loop $\beta$-function.}
\be
\alpha_s(\Lambda)&=&\frac{1}{b_0 t}\left[1-\frac{b_1}{b_0^2}\frac{\ln t}{t}+\frac{b_1^2(\ln^2 t-\ln t-1)+b_0b_2}{b_0^4 t^2}\right.
\nn
&&
\hspace{2cm}
\left.-\frac{b_1^3\lb\ln^3 t-\frac{5}{2}\ln^2 t-2\ln t+\frac{1}{2}\rb+3b_0b_1b_2\ln t}{b_0^6t^3}\right] ,
\ee
with $t = \ln(\Lambda^2/\Lambda_{\overline{\rm MS}}^2)$ and 
\be
b_0&=&\frac{11c_A-2N_f}{12\pi}\ ,\\
b_1&=&\frac{17c_A^2-5c_AN_f-2C_FN_f}{24\pi^2}\ ,\\
b_2&=&\frac{2857c_A^3+\lb54C_F^2-615C_F c_A-1415c_A^2\rb N_f+\lb66C_F+79c_A\rb N_f^2}{3456\pi^3} \, .
\ee
For one-loop running, we take $b_1=b_2=0$.  For both one- and three-loop running we fix the scale $\Lambda_{\overline{\rm MS}}$ by requiring that $\alpha_s({\rm 1.5\;GeV}) = 0.326$ which is obtained from lattice measurements \cite{latticealpha}.  For one-loop running, this procedure gives $\Lambda_{\overline{\rm MS}} = 176$ MeV, and for three-loop running, one obtains $\Lambda_{\overline{\rm MS}} = 316$ MeV.  

\subsection{Scales} 

For the renormalization scale we use separate scales, $\Lambda_g$ and $\Lambda_q$, for purely-gluonic and fermionic graphs, respectively.  We take the central values of these renormalization scales to be $\Lambda_g = 2\pi T$ and $\Lambda=\Lambda_q=2\pi \sqrt{T^2+\mu^2/\pi^2}$.  In all plots the thick lines indicate the result obtained using these central values and the light-blue band indicates the variation of the result under variation of both of these scales by a factor of two, e.g. $\pi T \leq \Lambda _g \leq 4 \pi T$.  For all numerical results below we use $c_A = N_c=3$ and $N_f=3$.

\subsection{Pressure} 

\begin{figure}[t]
\subfigure{
\hspace{-2mm}\includegraphics[width=7.5cm]{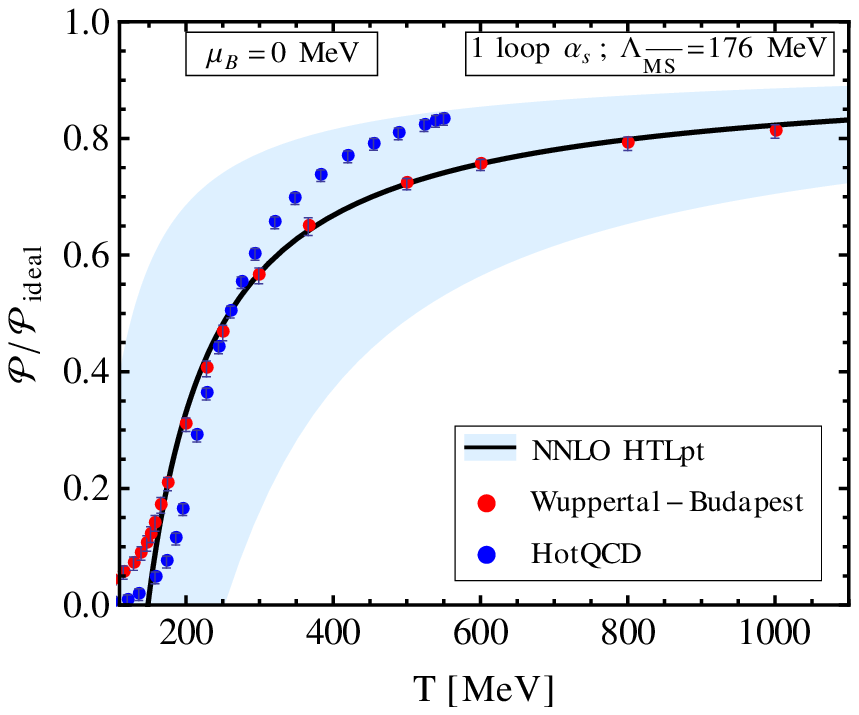}} 
\subfigure{
\includegraphics[width=7.5cm]{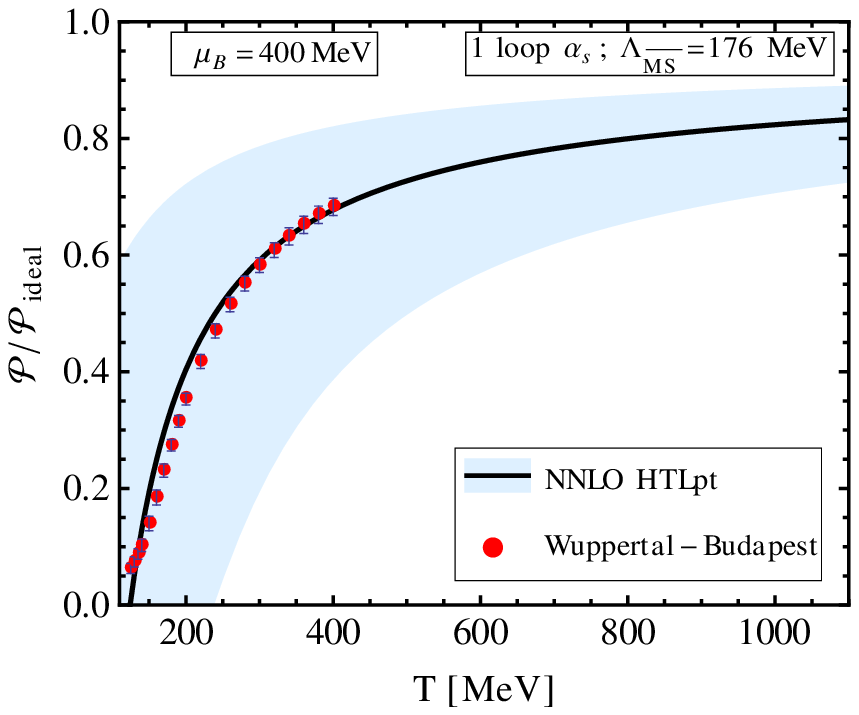}}
\caption{
Comparison of the $N_f=2+1$, $\mu_B=0$ (left) and $\mu_B=400$ MeV (right) NNLO HTLpt 
pressure with lattice data from Borsanyi et al. \cite{borsanyi1,borsanyi3} and Bazavov et al.~\cite{hotqcd1}.
For the HTLpt results a one-loop running coupling constant was used.
}
\label{pres_1l}
\end{figure}
\begin{figure}[t]
\subfigure{
\hspace{-2mm}\includegraphics[width=7.5cm]{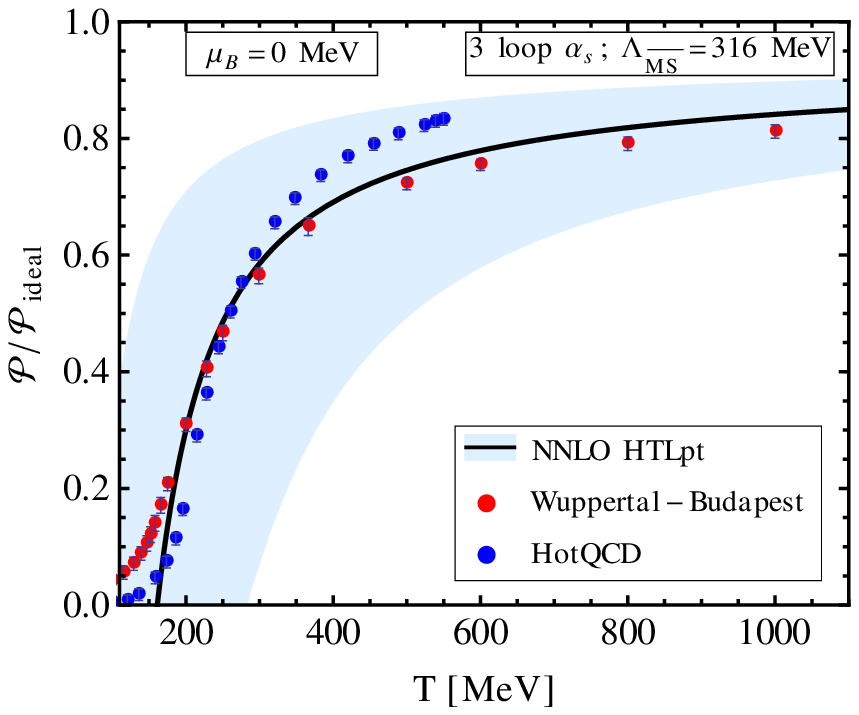}} 
\subfigure{
\includegraphics[width=7.5cm]{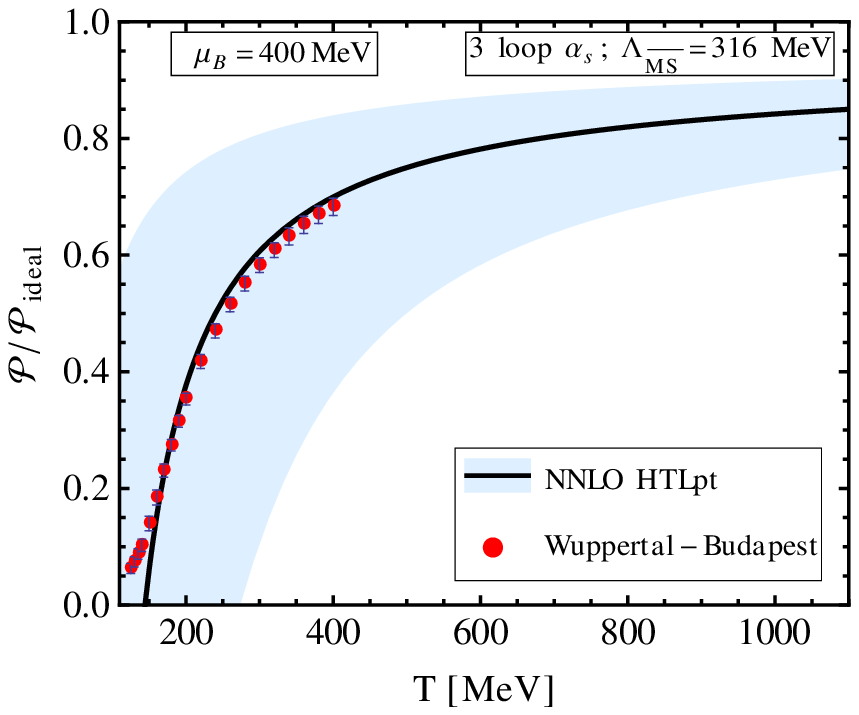}} 
\caption{
Same as fig.~\ref{pres_1l} except with a three-loop running coupling constant.
}
\label{pres_3l}
\end{figure}
\begin{figure}[t]
\centerline{\includegraphics[width=9cm]{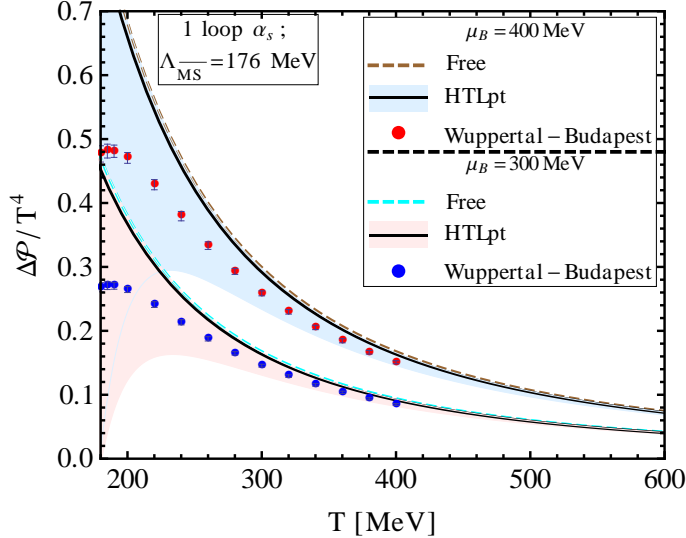}}
\caption{Comparison of the Stefan-Boltzmann limit (dashed lines) and NNLO HTLpt (solid lines) results for
the scaled pressure difference with lattice data from Borsanyi et al. \cite{borsanyi3}.}
\label{dPfig}
\end{figure} 

The QGP pressure can be obtained directly from the thermodynamic potential (\ref{finalomega1})
\be
{\cal P}(T,\Lambda,\mu)=-\Omega_{\rm NNLO}(T,\Lambda,\mu) \, ,
\ee
where $\Lambda$ above is understood to include both scales $\Lambda_g$ and $\Lambda_q$.
In figures~\ref{pres_1l} and \ref{pres_3l} we compare the scaled NNLO HTLpt pressure for $\mu_B=0$ (left) and $\mu_B=400$ MeV (right) with lattice data from refs.~\cite{hotqcd1,borsanyi1,Borsanyi:2012uq}.  In order to gauge the sensitivity of the results to the order of the running coupling, in fig.~\ref{pres_1l} we show the results obtained using a  one-loop running and in fig.~\ref{pres_3l} the results obtained using a three-loop running.  As can be seen by comparing these two sets, the sensitivity of the results to the order of the  running coupling is small for $T \gtrsim 250$ MeV.  As a result, unless the order of the running coupling turns out to have a significant effect on a given observable (see e.g. the fourth-order baryon number susceptibility), we will show the results obtained using a one-loop running coupling consistent with the counterterms necessary to renormalize the NNLO thermodynamic potential (\ref{ctalpha}).  

For an additional comparison we can compute the change in the pressure
\be
\Delta {\cal P} = {\cal P}(T,\Lambda,\mu)-{\cal P}(T,\Lambda,0) \, .
\ee
In figure~\ref{dPfig} we plot $\Delta {\cal P}$ as a function of the temperature for $\mu_B = 300$ MeV and $\mu_B = 400$ MeV.  The solid lines are the NNLO HTLpt result and the dashed lines are the result obtained in the Stefan-Boltzmann limit.  We note that  in fig.~\ref{dPfig} the lattice data from the Wuppertal-Budapest group  \cite{Borsanyi:2012uq} is computed up to ${\cal O}(\mu_B^2)$, whereas the HTLpt result includes all orders in $\mu_B$. As can be seen from this figure, the NNLO HTLpt result is quite close to the result obtained in the Stefan-Boltzmann limit.  Note that the small correction in going from the Stefan-Boltzmann limit to NNLO HTLpt indicates that the fermionic sector is, to good approximation, weakly coupled for $T \gtrsim 300$ MeV.

\subsection{Energy density}

\begin{figure}[t]
\subfigure{
\hspace{-2mm}\includegraphics[width=7.5cm]{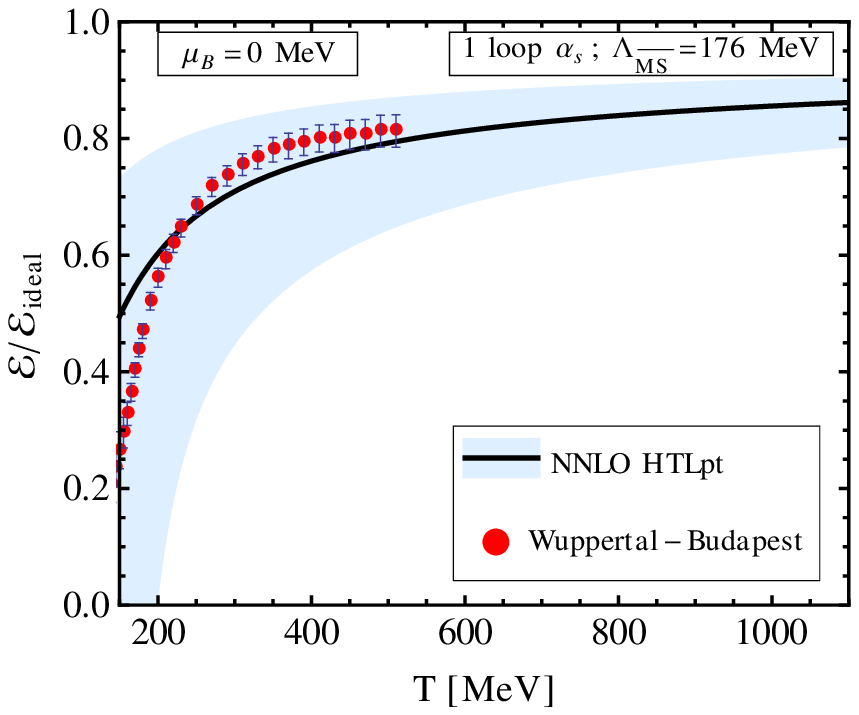}} 
\subfigure{
\includegraphics[width=7.5cm]{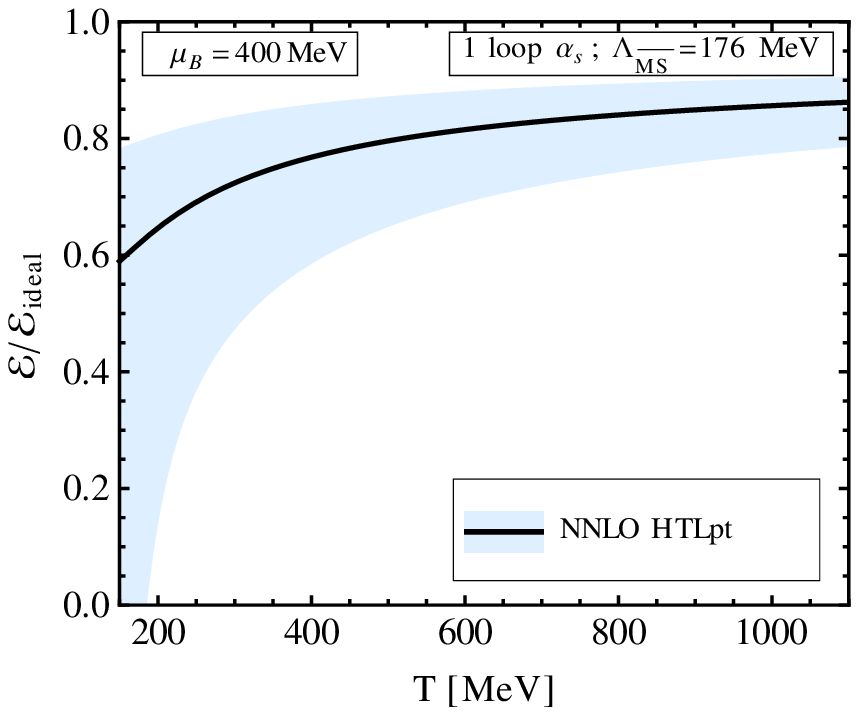}} 
\caption{
Comparison of the $N_f=2+1$, $\mu_B=0$ (left) and $\mu_B=400$ MeV (right) NNLO HTLpt 
energy density with lattice data.
The $\mu_B=0$ lattice data shown in the left panel are from ref.~\cite{borsanyi1}.
For the HTLpt results a one-loop running coupling constant was used.}
\label{ed_1l}
\end{figure}

Once the pressure is known, it is straightforward to compute other thermodynamic functions such
as the energy density by computing derivatives of the pressure with respect to the temperature
and chemical potential.  The energy density can be obtained via
\begin{eqnarray}
{\cal E}=T\frac{\partial{\cal P}}{\partial T}+\mu\frac{\partial{\cal P}}{\partial \mu}-{\cal P} \, .
\end{eqnarray}
In figure \ref{ed_1l} we plot the scaled NNLO HTLpt energy density for $\mu_B=0$ (left) and $\mu_B=400$ MeV (right) together with $\mu=0$ lattice data from ref.~\cite{borsanyi1}.  As we can see from this figure, there is reasonable agreement between the NNLO HTLpt energy density and the lattice data when the central value of the scale is used.

\subsection{Entropy density}

\begin{figure}[t]
\subfigure{
\hspace{-2mm}\includegraphics[width=7.5cm]{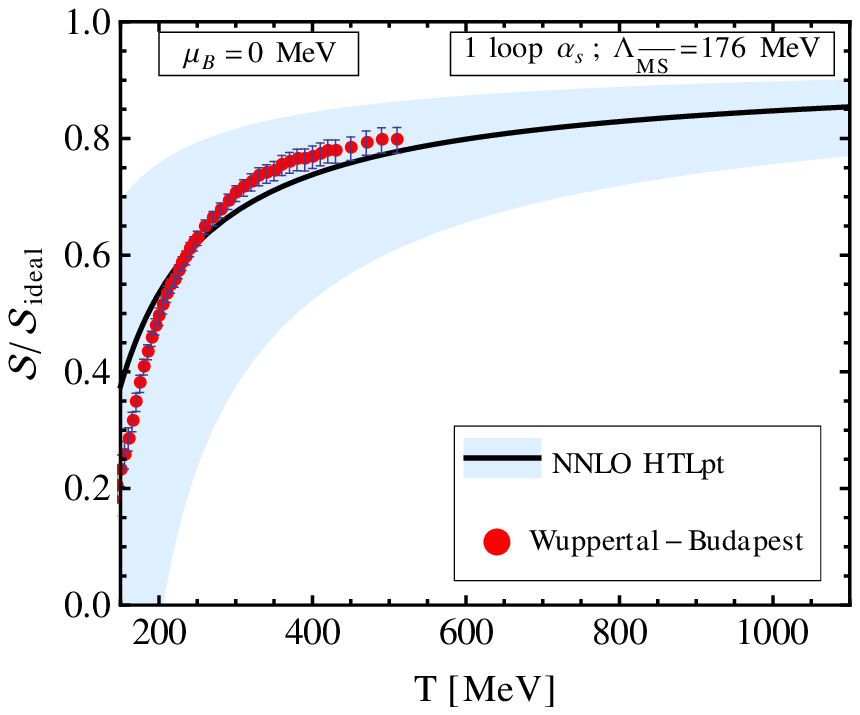}} 
\subfigure{
\includegraphics[width=7.5cm]{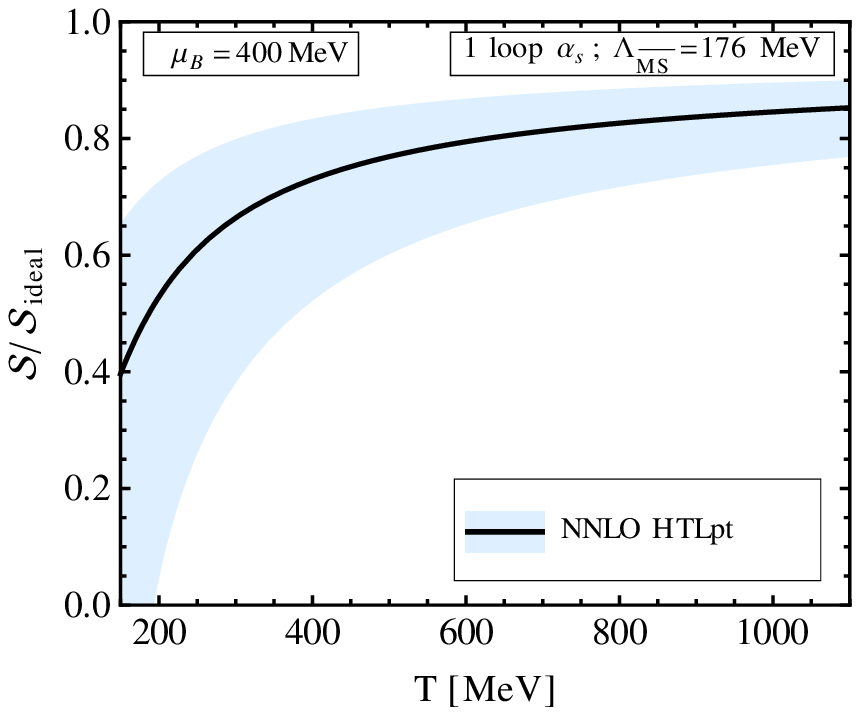}} 
\caption{
Comparison of the $N_f=2+1$, $\mu_B=0$ (left) and $\mu_B=400$ MeV (right) NNLO HTLpt 
entropy density with lattice data.
The $\mu_B=0$ lattice data shown in the left panel are from ref.~\cite{borsanyi1}.
For the HTLpt results a one-loop running coupling constant was used.}
\label{en_1l}
\end{figure}

Similarly, we can compute the entropy density
\be
{\cal S}(T,\mu) = \frac{\partial{\cal P}}{\partial T} \, .
\ee
We note that in the ideal gas limit, the entropy density becomes
\be
{\cal S}_{\rm ideal}(T,\mu)=\frac{4d_A\pi^2T^3}{45}\left[1+\frac{7}{4}\frac{d_F}{d_A}\left(1+\frac{60}{7}\hmu^2\right)\right] .
\ee
In figure \ref{en_1l} we plot the scaled NNLO HTLpt entropy density for $\mu_B=0$ (left) and $\mu_B=400$ MeV (right) together with $\mu=0$ lattice data from ref.~\cite{borsanyi1}.  As we can see from this figure, there is quite good agreement between the NNLO HTLpt entropy density and the lattice data when the central value of the scale is used.

\subsection{Trace anomaly}

\begin{figure}[t]
\subfigure{
\includegraphics[width=7.5cm]{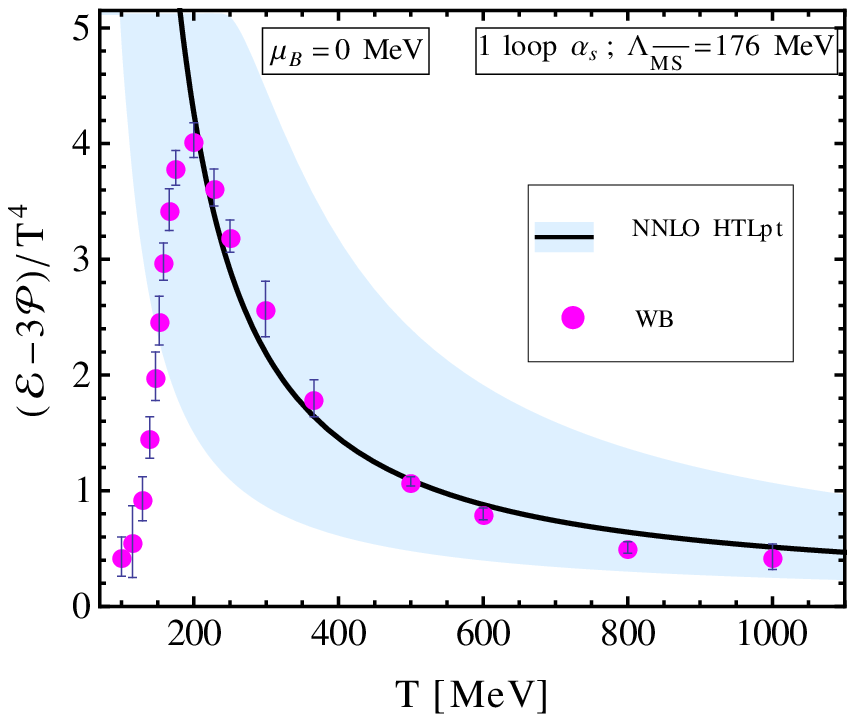}} 
\subfigure{
\includegraphics[width=7.5cm]{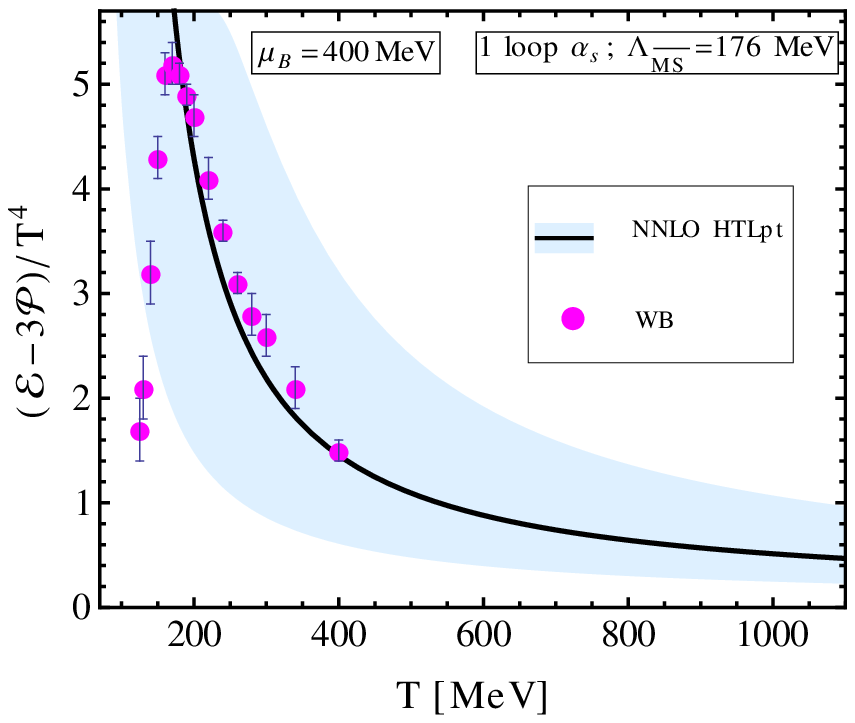}} 
\caption{
Comparison of the $N_f=2+1$, $\mu_B=0$ (left) and $\mu_B=400$ MeV (right) NNLO HTLpt 
trace anomaly with lattice data.
The $\mu_B=0$ lattice data are from \cite{borsanyi1} and the $\mu_B=400$ MeV lattice data are from \cite{borsanyi3}. 
For the HTLpt results a one-loop running coupling constant was used.
}
\label{ta_1l}
\end{figure}

Since it is typically the trace anomaly itself which is computed on the lattice and then integrated to obtain the other thermodynamic functions, it is interesting to compare directly with lattice data for the trace anomaly.  The trace anomaly is simply ${\cal I} = {\cal E}-3 {\cal P}$.  In the ideal gas limit, the trace anomaly goes to zero since ${\cal E}=3{\cal P}$.  When interactions are included, however, the trace anomaly (interaction measure) becomes non-zero.  In figure \ref{ta_1l} we plot the NNLO HTLpt trace anomaly scaled by $T^4$ for $\mu_B=0$ (left) and $\mu_B=400$ MeV (right) together with lattice data from refs.~\cite{borsanyi1} and \cite{borsanyi3}.  As we can see from this figure, there is quite good agreement between the NNLO HTLpt trace anomaly and the lattice data for $T \gtrsim 220$ MeV when the central value of the scale is used.

\subsection{Speed of sound}

\begin{figure}[t]
\subfigure{
\hspace{-4mm}\includegraphics[width=7.5cm]{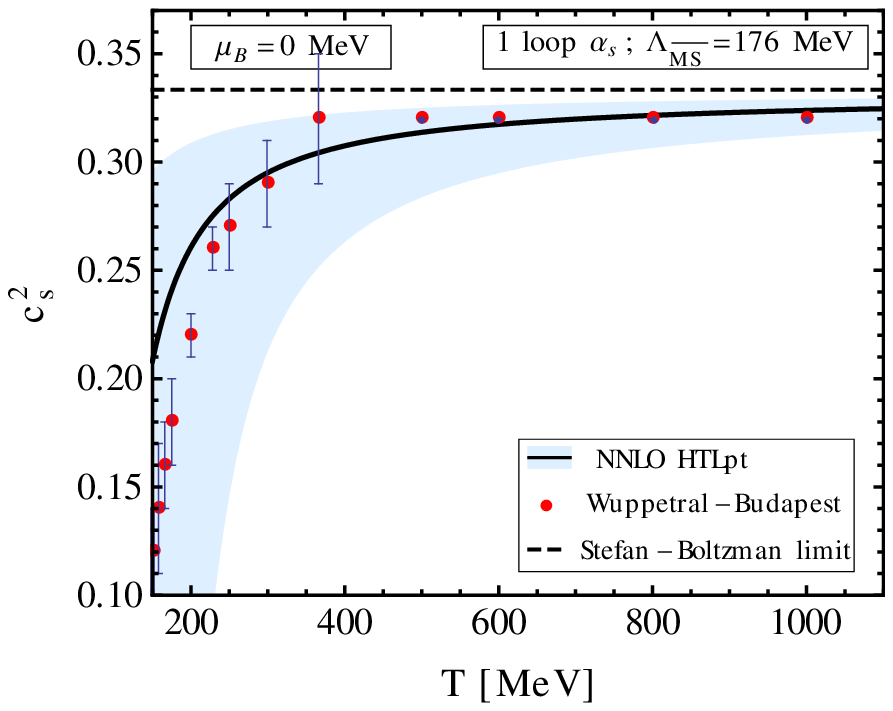}} 
\subfigure{
\includegraphics[width=7.5cm]{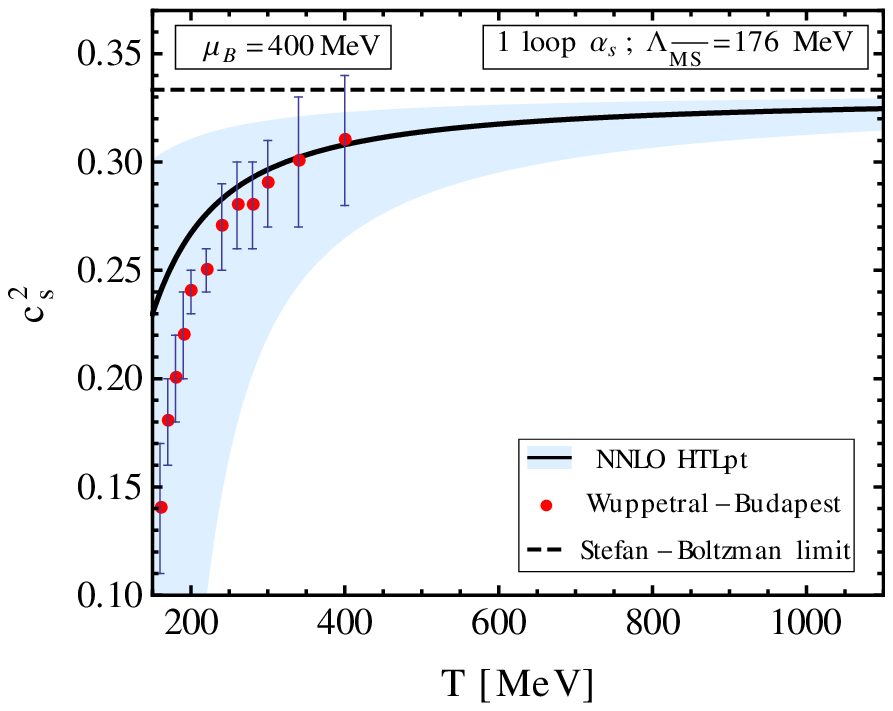}} 
\caption{
Comparison of the $N_f=2+1$, $\mu_B=0$ (left) and $\mu_B=400$ MeV (right) NNLO HTLpt 
speed of sound squared with lattice data.
The $\mu_B=0$ lattice data are from \cite{borsanyi1} and the $\mu_B=400$ MeV lattice data are from \cite{borsanyi3}. 
For the HTLpt results a one-loop running coupling constant was used.
}
\label{cssq_1l}
\end{figure}

Another quantity which is phenomenologically interesting is the speed of sound.  The speed of sound squared is defined as
\be
c_s^2=\frac{\del{\cal P}}{\del{\cal E}} \, .
\ee
In figure \ref{cssq_1l} we plot the NNLO HTLpt speed of sound for $\mu_B=0$ (left) and $\mu_B=400$ MeV (right) together with lattice data from refs.~\cite{borsanyi1} and \cite{borsanyi3}.  As we can see from this figure, there is quite good agreement between the NNLO HTLpt speed of sound and the lattice data when the central value of the scale is used.

\section{Quark number susceptibilities}
\label{sec:qns}

Using the full thermodynamic potential as a function of chemical potential(s) and temperature we can compute the quark number susceptibilities.  In general, one can introduce a separate chemical potential for each quark flavor giving a $N_f$-dimensional vector $\bm{\mu}\equiv(\mu_u,\mu_d,...,\mu_{N_f})$.  By taking derivatives of the pressure with respect to chemical potentials in this set, we obtain the quark number susceptibilities\,\footnote{We have specified that the derivatives should be evaluated at $\bm{\mu}=0$. In general, one could define the susceptibilities at $\bm{\mu} = \bm{\mu}_0$.}
\be
\chi_{ijk\,\cdots}\left(T\right)&\equiv& \left. \frac{\partial^{i+j+k+ \, \cdots}\; {\cal P}\left(T,\bm{\mu}\right)}{\partial\mu_u^i\, \partial\mu_d^j \, \partial\mu_s^k\, \cdots} \right|_{\bm{\mu}=0} \, .
\label{qnsdef}
\ee
Below we will use a shorthand notation for the susceptibilities by specifying derivatives by a string of quark flavors in superscript form, e.g. $\chi^{uu}_2 = \chi_{200}$, $\chi^{ds}_2 = \chi_{011}$, $\chi^{uudd}_4 
= \chi_{220}$, etc.

When computing the derivatives with respect to the chemical potentials we treat $\Lambda_q$ as being a constant and only put the chemical potential dependence of the $\Lambda_q$ in after the derivatives are taken.  We have done this in order to more closely match the procedure used to compute the susceptibilities using resummed dimensional reduction \cite{sylvain2}.\footnote{One could instead put the chemical potential dependence of the $\Lambda_q$ in prior to taking the derivatives with respect to the chemical potentials.  If this is done, the central lines obtained are very close to the ones obtained using the fixed-$\Lambda_q$ prescription, however, the scale variation typically increases in this case.}

\subsection{Baryon number susceptibilities}

\begin{figure}[t]
\subfigure{
\includegraphics[width=7.5cm]{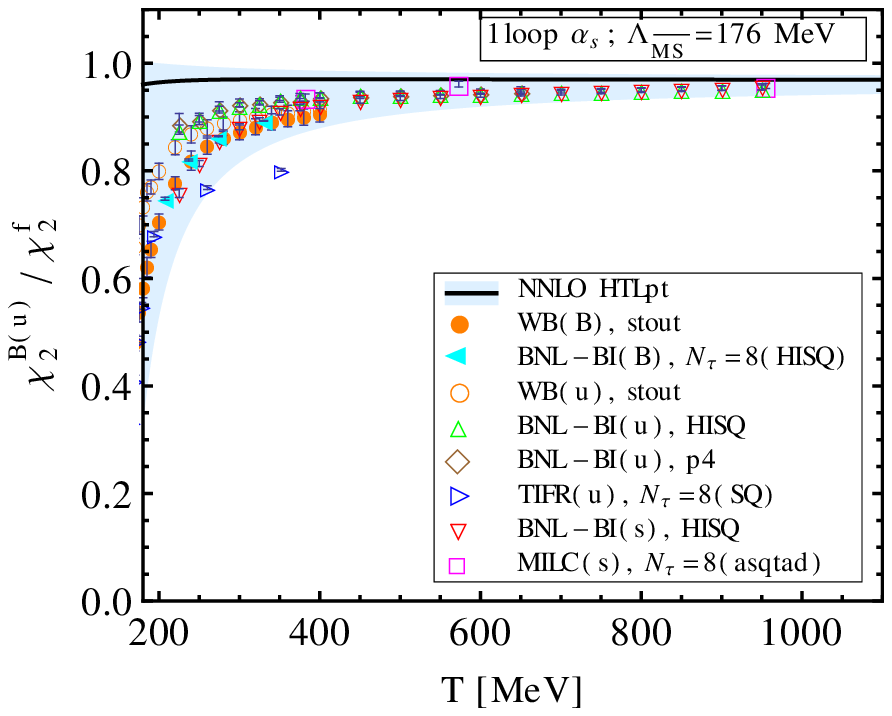}}
\subfigure{
\includegraphics[width=7.5cm]{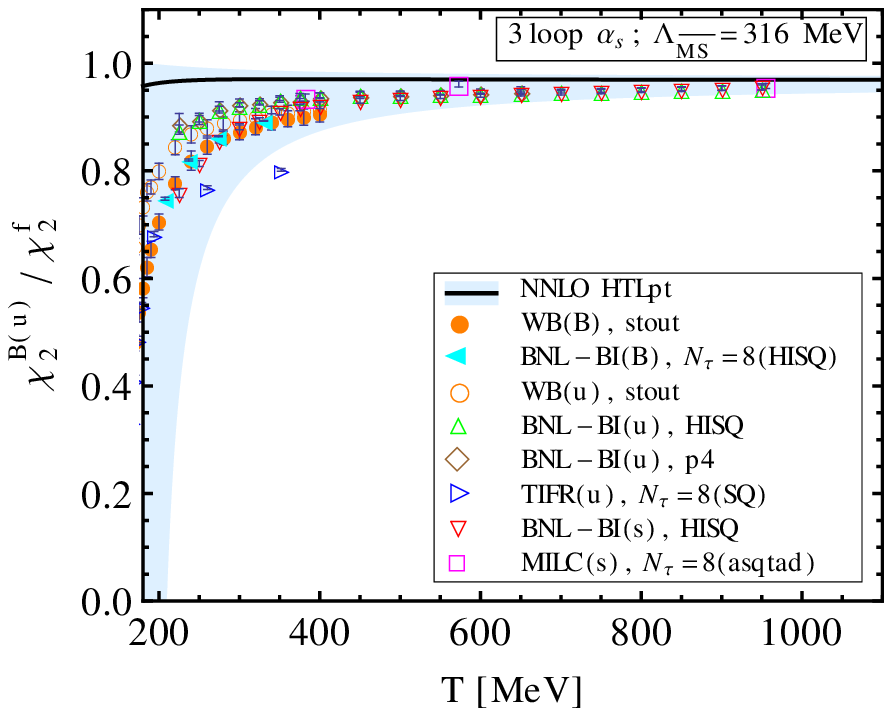}}
\caption{
The scaled second order baryon number susceptibility compared with various lattice data using one-loop running (left) and three-loop running (right). The lattice data labeled WB, BNL-BI(B), BNL-BI(u,s), MILC, and TIFR come from refs.~\cite{borsanyi2}, \cite{bnlb1}, \cite{bnlb2}, \cite{milc}, and \cite{TIFR}, respectively.
}
\label{qns2_1l}
\end{figure}
\begin{figure}[t]
\subfigure{
\includegraphics[width=7.5cm]{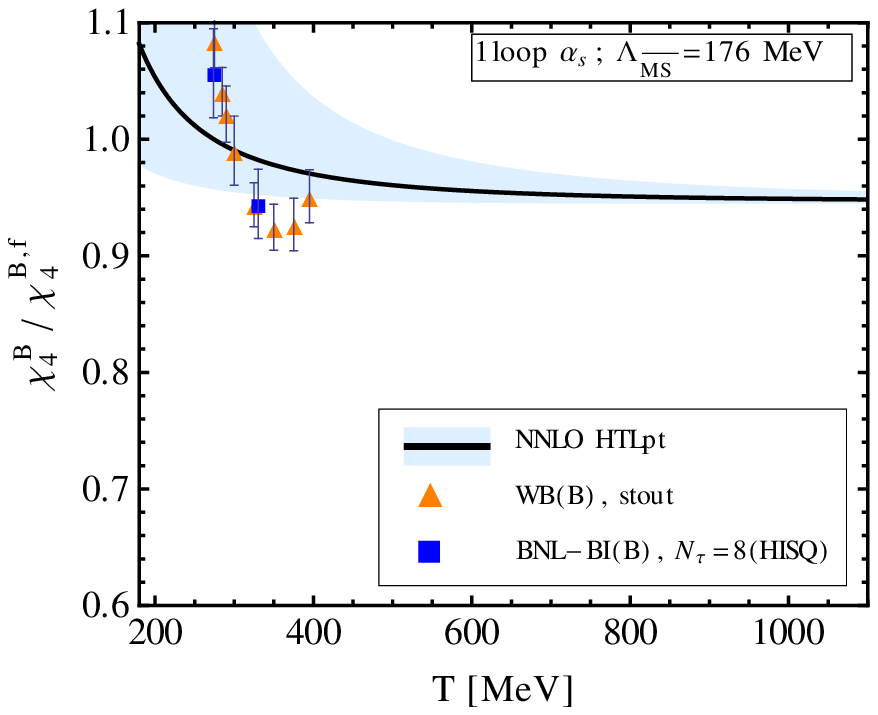}}
\subfigure{
\includegraphics[width=7.5cm]{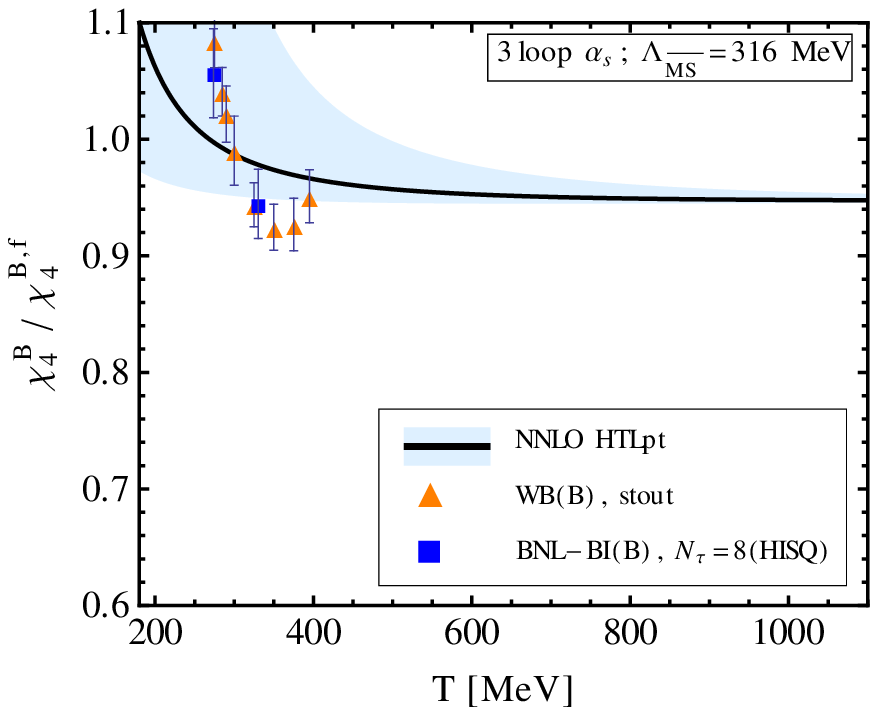}}
\caption{
The scaled fourth order baryon number susceptibility compared with various lattice data using one-loop running (left) and three-loop running (right). The lattice data labeled WB, BNL-BI(B), BNL-BI(u,s), MILC, and TIFR come from refs.~\cite{borsanyi2}, \cite{bnlb1}, \cite{bnlb2}, \cite{milc}, and \cite{TIFR}, respectively.
}
\label{qns4_1l}
\end{figure}
\begin{figure}[t]
\centerline{\includegraphics[width=7.5cm]{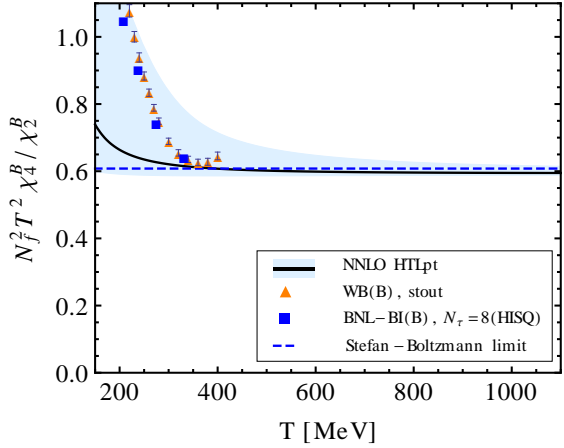}}
\caption{
Comparison of the $N_f=2+1$ NNLO HTLpt ratio of the fourth to second order baryon susceptibility with lattice data.  For the HTLpt results a one-loop running coupling constant was used.  The data labeled WB and BNL-BI(B) come from refs.~\cite{borsanyi4,borsanyi5} and \cite{bnlb1}, respectively.
}
\label{qnsrat_1l}
\end{figure}

We begin by considering the baryon number susceptibilities.  The $n^{\rm th}$-order baryon number susceptibility is defined as
\be
\chi_B^n(T) \equiv \left.\frac{\partial^n {\cal P}}{\partial \mu_B^n}\right|_{\mu_B=0} \, .
\ee
For a three flavor system consisting of $(u,d,s)$, the baryon number susceptibilities can be related to the quark number susceptibilities~\cite{peter_review}
\be
\chi_2^B=\frac{1}{9}\[\chi_2^{uu}+\chi_2^{dd}+\chi_2^{ss}+2\chi_2^{ud}+2\chi_2^{ds}+2\chi_2^{us}\] \, ,
\label{gen_chi2}
\ee
and
\be
\chi_4^B &=& \frac{1}{81}\[\chi_4^{uuuu}+\chi_4^{dddd}+\chi_4^{ssss}+4\chi_4^{uuud}+4\chi_4^{uuus}+4\chi_4^{dddu}+4\chi_4^{ddds}+
4\chi_4^{sssu}\right.\nn
&&
\hspace{8mm}
\left.+ \, 4\chi_4^{sssd}+6\chi_4^{uudd}+6\chi_4^{ddss}+6\chi_4^{uuss}+12\chi_4^{uuds}+12\chi_4^{ddus}+12\chi_4^{ssud} \] \, .
\hspace{5mm}
\label{gen_chi4}
\ee
If we treat all quarks as having the same chemical potential $\mu_u=\mu_d=\mu_s=\mu=\frac{1}{3}\mu_B$, eqs.~(\ref{gen_chi2}) and (\ref{gen_chi4}) reduce to $\chi_2^B=\chi_2^{uu}$ and $\chi_4^B=\chi_4^{uuuu}$.  This allows us to straightforwardly compute the baryon number susceptibility by computing derivatives of (\ref{finalomega1}) with respect to $\mu$.

\begin{figure}[t]
\centerline{\includegraphics[width=7.5cm]{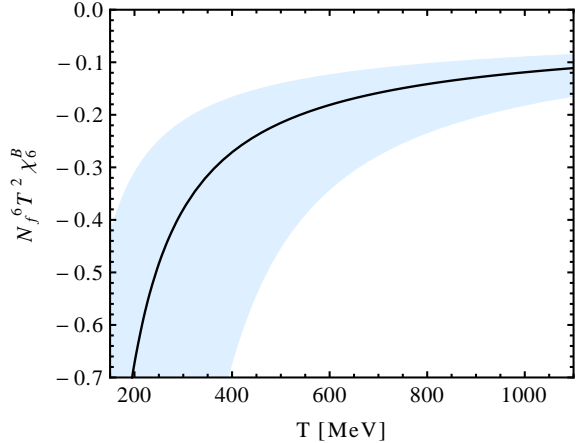}}
\caption{
The $N_f=2+1$ NNLO HTLpt scaled sixth-order baryon susceptibility as a function of temperature.
}
\label{qns6_1l}
\end{figure}

In figure~\ref{qns2_1l} we compare the NNLO HTLpt result for the second order baryon number susceptibility with lattice data from various groups.  In the left panel of this figure we used the one-loop running and on the right we used the three-loop running.  As one can see, for this quantity, the size of the light-blue band becomes larger if one uses the three-loop running, however, the central value obtained is very close in both cases.  

Comparing to the lattice data we see that the NNLO HTLpt prediction is approximately 10\% higher than the lattice data at $T=250$ MeV and approximately 2\% higher at $T = 800$ MeV.  We note in this context that recently the four-loop second-order baryon number susceptibility has been computed in ref.~\cite{sylvain2} using the resummed dimensional reduction method.  The result from this approach lies within the NNLO HTLpt scale variation band and is even closer to the lattice data with the error at $T=250$ MeV being approximately 2\% and $\lesssim 1\%$ at  $T = 800$ MeV.  Our result, taken together with the resummed dimensional reduction results seem to indicate that the quark sector of the QGP can be quite accurately described using resummed perturbation theory for temperatures above approximately 300 MeV.

In figure~\ref{qns4_1l} we compare the NNLO HTLpt result for the fourth order baryon number susceptibility with lattice data.  Once again we show in the left and right panels, the result obtain using the one-loop running coupling and three-loop running coupling, respectively.  Both the one- and three-loop running results are consistent with the lattice data shown; however, the lattice error bars on this quantity are somewhat large and the data are restricted to temperatures below 400 MeV, making it difficult to draw firm conclusions from this comparison.  That being said, HTLpt makes a clear prediction for the temperature dependence of the fourth order baryon number susceptibility.  It will be very interesting to see if future lattice data agree with this prediction.

In figure~\ref{qnsrat_1l} we plot the scaled ratio of the fourth and second order baryon number susceptibilities as a function of temperature along with lattice data for this ratio.  As we can see from
this figure, this ratio very rapidly approaches the Stefan-Boltzmann limit if one considers the central NNLO HTLpt line.  Comparing with the lattice data we see that the NNLO HTLpt result is below the lattice data for temperatures less than approximately 300 MeV.  Without lattice data at higher temperatures, it is hard to draw a firm conclusion regarding the temperature at which HTLpt provides a good description of this quantity.

In figure~\ref{qns6_1l} we show the NNLO HTLpt prediction for the sixth order baryon number susceptibility.  To the best of our knowledge there is currently no publicly available lattice data for this quantity.  It will be very interesting to see if these NNLO HTLpt predictions agree with lattice data as they becomes available.

\subsection{Single quark number susceptibilities}

\begin{figure}[t]
\subfigure{
 \includegraphics[width=7.5cm]{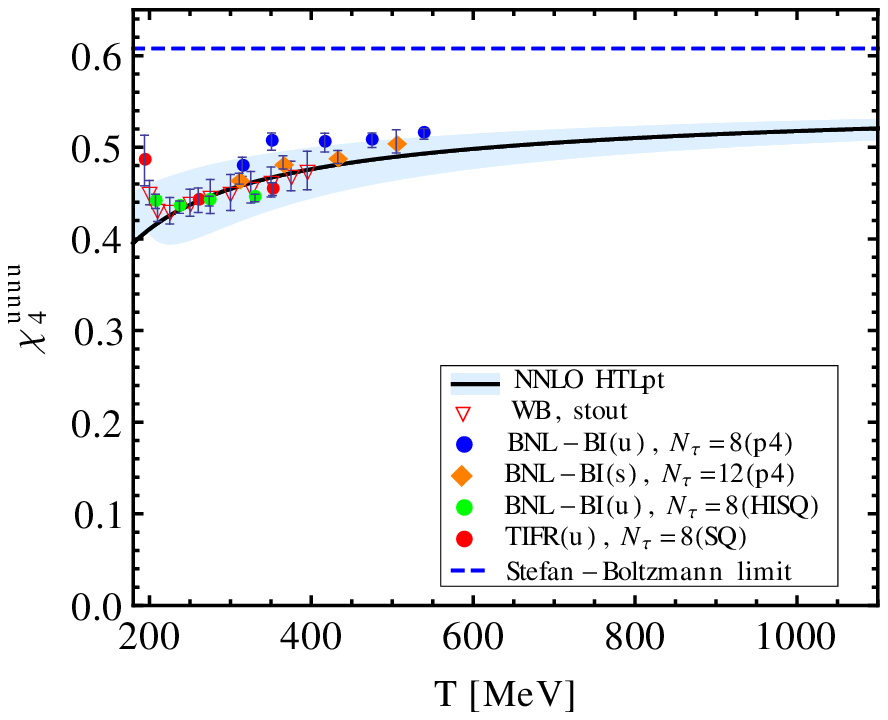}}
\subfigure{
\includegraphics[width=7.5cm]{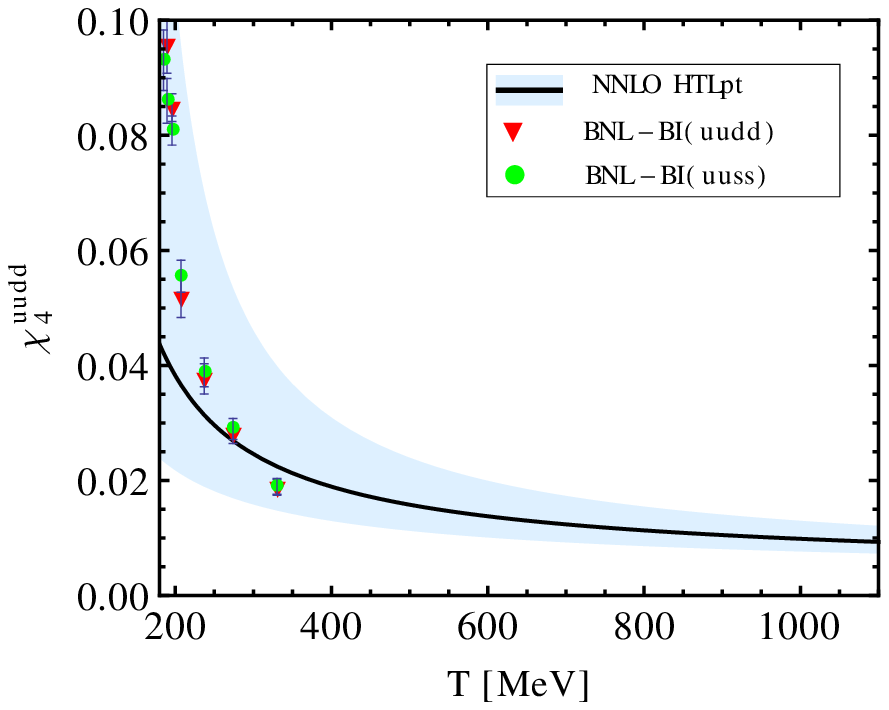}}
\caption{
Comparison of the $N_f=2+1$ NNLO HTLpt ratio of the fourth order diagonal single quark number susceptibility (left) and the only non-vanishing fourth order off-diagonal quark number susceptibility (right) with lattice data.  In the left figure the dashed blue line indicates the Stefan-Boltzmann limit for this quantity.  For the HTLpt results a one-loop running coupling constant was used.  The data labeled BNL-BI(uudd), BNL-BI(u,s), BNL-BI(uuss), and TIFR come from refs.~\cite{bnlb1}, \cite{bnlb2}, \cite{bnlb3}, and \cite{TIFR}, respectively.
}
\label{figsingleq}
\end{figure}
\begin{figure}[t]
\vspace{3mm}
\centerline{\includegraphics[width=7.5cm]{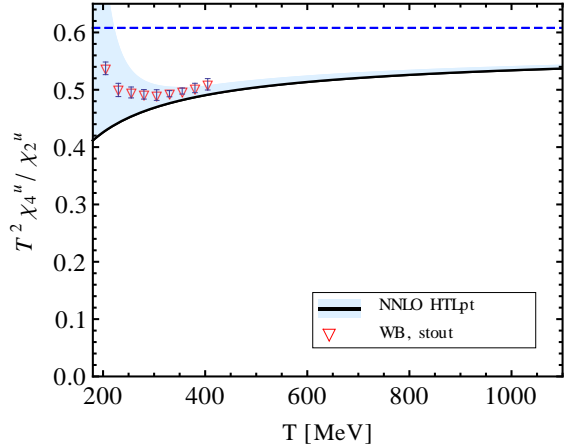}}
\caption{
Comparison of the $N_f=2+1$ NNLO HTLpt ratio of the fourth to second order single quark susceptibility with lattice data.  For the HTLpt results a one-loop running coupling constant was used.  The data labeled WB come from refs.~\cite{borsanyi4,borsanyi5}.}
\label{qnsrat2_1l}
\end{figure}

We now consider the single quark number susceptibilities (\ref{qnsdef}).  For these we use the general expression for the NNLO thermodynamic potential with different quark chemical potentials (\ref{finalomega}).  The resulting susceptibilities can either be diagonal (same flavor on all derivatives) or off-diagonal (different flavor on some or all indices).  In HTLpt there are off-diagonal susceptibilities emerging explicitly from graphs ${\cal F}_{3c}^f$ and ${\cal F}_{3j}^f$; however, the latter  vanishes when we use the variational mass prescription for the quark mass ($m_q=0$), so we need only consider the ${\cal F}_{3c}^f$ graph.  Additionally, there are potential off-diagonal contributions coming from all HTL terms since the Debye mass receives contributions from all quark flavors.  In practice, however, because we evaluate derivatives with respect to the various chemical potentials and then take $\mu_f \rightarrow 0$, one finds that all off-diagonal second order susceptibilities vanish in HTLpt.  Therefore, for the three-flavor case one has
\be
\chi_2^{ud}=\chi_2^{ds}=\chi_2^{su}=0 \, ,
\ee
and, as a result, the single quark second order susceptibility is proportional to the baryon number susceptibility
\be
\chi_2^{uu}=\frac{1}{3}\chi_2^B.
\ee
For the fourth order susceptibility, there is only one non-zero off-diagonal susceptibility, namely $\chi_4^{uudd}=\chi_4^{uuss}=\chi_4^{ddss}$, which is related to the diagonal susceptibility, e.g. $\chi_4^{uuuu}=\chi_4^{dddd}=\chi_4^{ssss}$, as 
\be
\chi_4^{uuuu}=27\chi_4^B-6\chi_4^{uudd}.
\label{u4rel}
\ee
As a consequence, one can compute $\chi_4^{uuuu}$ directly from (\ref{finalomega}) or by computing $\chi_4^B$ using (\ref{finalomega1}) and $\chi_4^{uudd}$ using (\ref{finalomega}) and applying the above relation.  In our final plots we compute $\chi_4^{uuuu}$ directly from (\ref{finalomega}), however, we have checked that we obtain the same result if we use (\ref{u4rel}) instead.

In figure~\ref{figsingleq} (left) we plot our result for the fourth order single quark susceptibility $\chi_4^{uuuu}$ compared to lattice data from refs.~\cite{bnlb2}, \cite{bnlb1}, \cite{bnlb3}, and \cite{TIFR}.  As we can see from this figure, for the fourth order susceptibility there is very good agreement with available lattice data.  In addition, the scale variation of the HTLpt result is quite small for this particular quantity.  In figure~\ref{figsingleq} (right) we plot our result for the fourth order off-diagonal single quark susceptibility $\chi_4^{uudd}$ compared to lattice data.  From this right panel we also see reasonably good agreement between the NNLO HTLpt result and the available lattice data.  

In figure~\ref{qnsrat2_1l} we plot the scaled ratio of the fourth- and second-order single quark susceptibilities.  Once again we see good agreement between the NNLO HTLpt result and lattice data.  Once again, for both figures~\ref{figsingleq} and \ref{qnsrat2_1l}, the lattice data are confined to relatively low temperatures. It will be interesting to compare higher temperature lattice data with the NNLO HTLpt prediction as they become available.   

\begin{figure}[t]
\includegraphics[width=4.9cm]{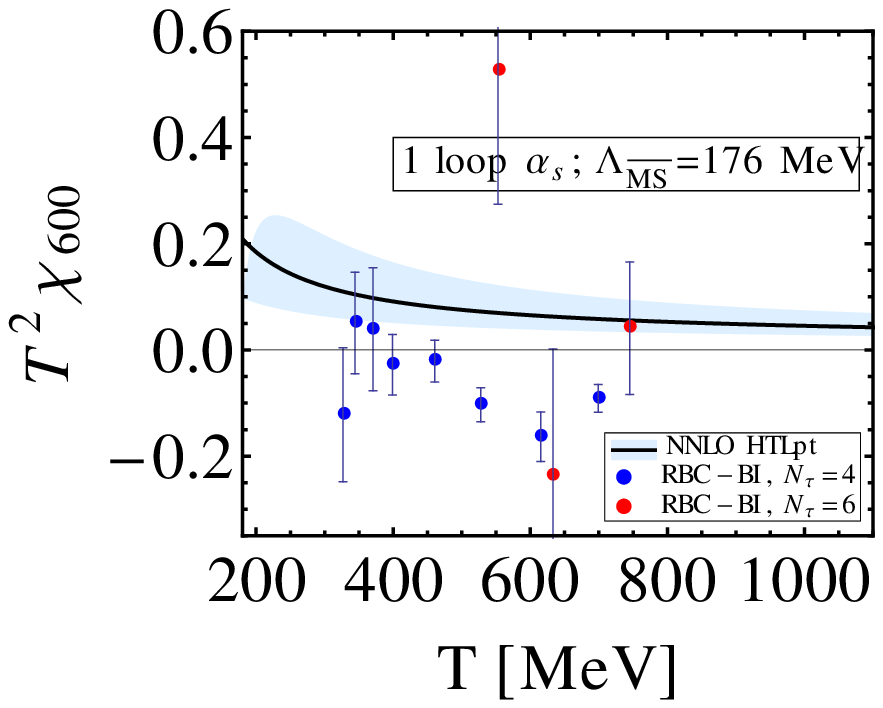}
\includegraphics[width=4.9cm]{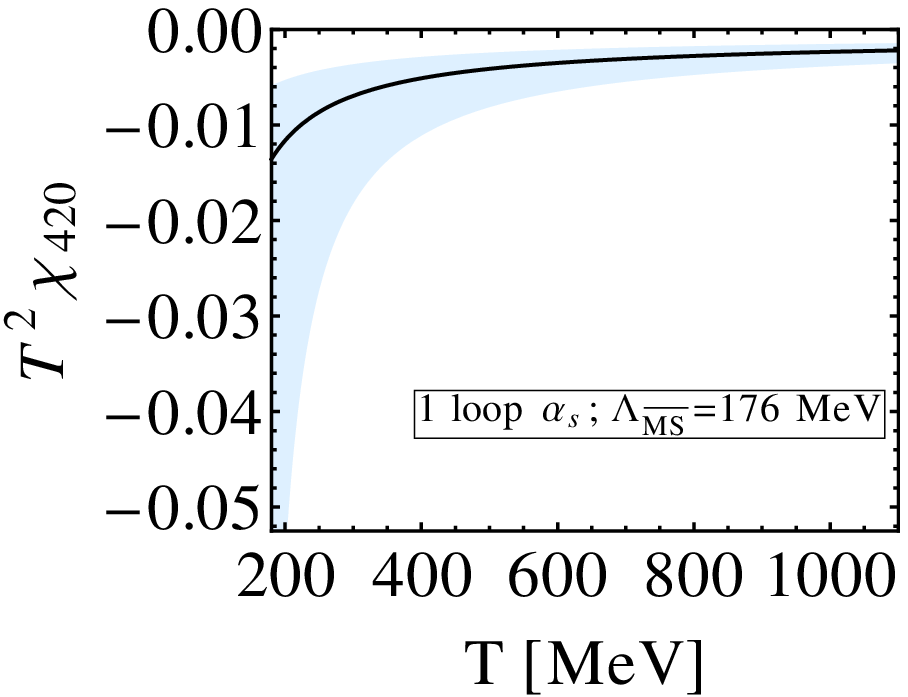}
\includegraphics[width=4.9cm]{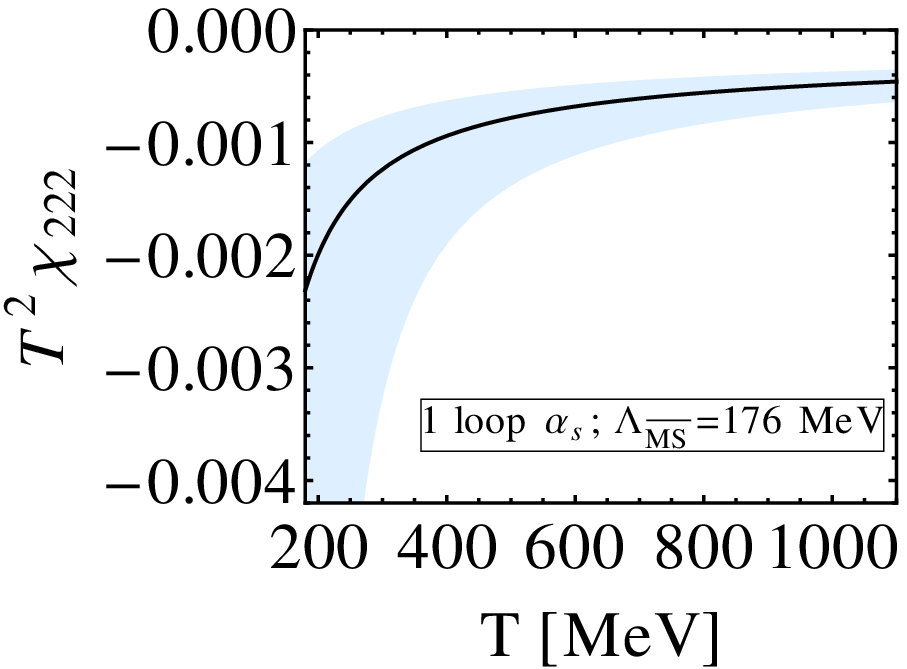}
\caption{
The $N_f=2+1$ NNLO HTLpt scaled sixth-order diagonal and off-diagonal single quark susceptibilities $\chi_{600}$ (left), $\chi_{420}$ (middle), and $\chi_{222}$ (right) as a function of temperature.  In the left panel we show lattice data available from the RBC-Bielefeld collaboration for $N_\tau = 4$ and $N_\tau = 6$ lattices~\cite{Petreczky:2009cr}.
}
\label{qns6single_1l}
\end{figure}

Finally, in figure~\ref{qns6single_1l} we plot the diagonal and off-diagonal sixth-order quark number susceptibilities 
$\chi_{600}$ (left), $\chi_{420}$ (middle), and $\chi_{222}$ (right).
In the left panel of figure~\ref{qns6single_1l} we show lattice data available from the RBC-Bielefeld collaboration \cite{Petreczky:2009cr}.  At this point in time the lattice sizes are small and the errors bars for $\chi_{600}$ are large, so it is hard to draw a firm conclusion from this comparison.  Regarding the off-diagonal six-order single quark susceptibilities (center and right panels), we are unaware of any lattice data for these.
As before, it will be very interesting to see if these NNLO HTLpt predictions agree with lattice data as it becomes available.

\section{Conclusions and outlook}
\label{outlook}

In this paper, we presented the results of a NNLO (three-loop) HTLpt calculation of the thermodynamic potential of QCD at finite temperature and chemical potential(s).  Our final result (\ref{finalomega}) is completely analytic, gauge invariant, and should be valid in the region of the phase diagram for which $\mu_f \lesssim 2 \pi T$.  Based on the resulting thermodynamic potential we proceeded to calculate the pressure, energy density, entropy density, trace anomaly, and speed of sound of the QGP.  In all cases we found very good agreement between the results obtained using the central values of the renormalization scales and available lattice data.  Additionally, we have made predictions for the diagonal and off-diagonal sixth-order baryon number and single quark susceptibilities.

Looking to the future there are still many avenues for improvement in the HTLpt approach:  (1) inclusion of the effects of finite quark masses (2) extension of results to $\mu_f \gtrsim 2 \pi T$ and eventually to $T=0$, and (3) to potentially resum logarithms in order to reduce the scale variation of the final results (light-blue bands in all figures).  Of these three, the second task is the most straightforward; however, in order to make more definitive and constrained statements it now seems necessary to start moving in directions (1) and (3) as well.  In closing, we emphasize that HTLpt provides a gauge invariant reorganization of perturbation theory for calculating static quantities in thermal field theory.  Since the NNLO HTLpt results are in good agreement with lattice data for various thermodynamic quantities down to temperatures that are relevant for LHC, it would therefore be interesting and challenging to apply HTLpt to the calculation of dynamic quantities, especially transport coefficients, at these temperatures.

\acknowledgments{ 
We thank S. Borsanyi, S. Datta, F. Karsch, S. Gupta, S. Mogliacci, P. Petreczky, and A. Vuorinen for useful discussions. N.H., A.B., and M.G.M. were supported by the Indian Department of Atomic Energy. M.S. was supported in part by DOE Grant No.~\mbox{DE-SC0004104}. N.S. was supported by the Bielefeld Young Researchers' Fund. 
}


\appendix

\section{Expansion in mass parameters}
\label{expansion}

In refs.~\cite{andersen4,andersen5} the NLO HTLpt thermodynamic potential was reduced to scalar sum-integrals.  Evaluating these scalar sum-integrals exactly seems intractable, however, the sum-integrals can be calculated approximately by expanding them in powers of $m_D/T$ and $m_q/T$ following the method developed in ref.~\cite{spt5}. We will adopt the same strategy in this paper and include all terms through order $g^5$ assuming that $m_D$ and $m_q$ are ${\cal O}(g)$ at leading order.  At each loop order, the contributions can be divided into those coming from hard and soft momenta, which are the momenta proportional to the scales $T$ and $gT$, respectively. In the one-loop diagrams, the contributions are either hard $(h)$ or soft $(s)$, while at the two-loop level, there are hard-hard $(hh)$, hard-soft $(hs)$, and soft-soft $(ss)$ contributions. At three loops there are hard-hard-hard $(hhh)$, hard-hard-soft $(hhs)$, hard-soft-soft $(hss)$, and soft-soft-soft $(sss)$ contributions.

\subsection{One-loop sum-integrals}

We now review the mass expansion of the necessary one-loop sum-integrals considering separately
the contributions from hard and soft momenta.  We list the purely gluonic contributions
when they are necessary for simpler exposition of the final result.  Note that in order to simplify
the results, when possible, it is best to add the corresponding iterated polarization and self-energy 
insertions that appear at higher order in $\delta$, e.g. below we will also include ${\cal F}^g_{\rm 2d}$,
${\cal F}^f_{\rm 2d}$, ${\cal F}^g_{\rm 3m}$, and ${\cal F}^f_{\rm 3i}$ as ``one-loop'' contributions.

\subsubsection*{Hard contributions}

For one-loop gluon $({\cal F}_{1a}^g)$ and one-loop ghost $({\cal F}_{1b}^g)$ diagrams, we need to expand in order $m_D^2$:
\be
{\cal F}_{\rm 1a+1b}^{g(h)}&=&\frac{1}{2}(d-1)\sumintb_P\ \ln P^2+\frac{1}{2}
m_D^2\sumintb_P\frac{1}{P^2}
\nn
&&
-\frac{1}{4(d-1)}m_D^4\sumintb_P\left[\frac{1}{P^4}
-\frac{2}{p^2P^2}-\frac{2d}{p^4}{\cal T}_P
+\frac{2}{p^2P^2}{\cal T}_P
+\frac{d}{p^4}{\cal T}_P^2
\right].
\label{Flo-h}
\ee
The one-loop graph with a gluon self-energy insertion $({\cal F}_{2d}^g)$ has an explicit factor of $m_D^2$ and, therefore, we only need to expand the sum-integral to first order in $m_D^2$:
\be
{\cal F}_{\rm 2d}^{g(h)}\!\!&=&\!\!-\frac{1}{2}
m_D^2\sumintb_P\frac{1}{P^2}
+\frac{1}{2(d-1)}m_D^4\sumintb_P\left[
\frac{1}{P^4}-\frac{2}{p^2P^2}-\frac{2d}{p^4}{\cal T}_P
+\frac{2}{p^2P^2}{\cal T}_P
+\frac{d}{p^4}{\cal T}_P^2
\right]. \hspace{6mm}
\label{ct1}
\ee
The one-loop graph with two gluon self-energy insertions $({\cal F}_{3m}^g)$ 
must be expanded to zeroth order in $m_D^2$
\be
{\cal F}_{\rm 3m}^{g(h)}&=&
-\frac{1}{4(d-1)}m_D^4\sumintb_P\left[
\frac{1}{P^4}-\frac{2}{p^2P^2}-\frac{2d}{p^4}{\cal T}_P
+\frac{2}{p^2P^2}{\cal T}_P
+\frac{d}{p^4}{\cal T}_P^2
\right] .
\label{ct2}
\ee
The sum of eqs.~(\ref{Flo-h})-(\ref{ct2}) is very simple
\be
{\cal F}_{\rm 1a+1b+2d+3m}^{g(h)} =
\frac{1}{2}(d-1)\sumintb_P\ \ln\left(P^2\right) = -\frac{\pi^2}{45}T^4
\;.
\ee

The one-loop fermionic graph ${\cal F}_{1b}^f$ needs to expanded to second order in $m^2_q$
\be
{\cal F}_{\rm 1b}^{f(h)}&=&-2\sumintf_{\{P\}}\log P^2-4m_q^2
\sumintf_{\{P\}}\frac{1}{P^2}
+2m_q^4\sumintf_{\{P\}}\left[
\frac{2}{P^4}
-\frac{1}{p^2P^2}+
\frac{2}{p^2P^2}{\cal T}_P
-\frac{1}{p^2P_0^2}{\cal T}_P^2
\right] \! . \hspace{9mm}
\label{f1b}
\ee
The one-loop fermion loop with a fermion self-energy insertion ${\cal F}_{2d}^f$
must be expanded to first order in $m_q^2$,
\be
{\cal F}_{\rm 2d}^{f(h)}&=&4m_q^2\sumintf_{\{P\}}\frac{1}{P^2}
-4m_q^4\sumintf_{\{P\}}\left[
\frac{2}{P^4}-\frac{1}{p^2P^2}
+
\frac{2}{p^2P^2}{\cal T}_P
-\frac{1}{p^2P_0^2}{\cal T}_P^2
\right].
\label{f2d}
\ee
The one-loop fermion loop with two self-energy insertions ${\cal F}_{3i}^f$ must be
expanded to zeroth order in $m_q^2$:
\be
{\cal F}_{\rm 3i}^{f(h)}&=&2m_q^4\sumintf_{\{P\}}\left[
\frac{2}{P^4}-\frac{1}{p^2P^2}
+
\frac{2}{p^2P^2}{\cal T}_P
-\frac{1}{p^2P_0^2}{\cal T}_P^2
\right].
\label{f3i}
\ee
The sum of eqs.~(\ref{f1b})-(\ref{f3i}) is particularly simple
\be
{\cal F}_{\rm 1b+2d+3i}^{f(h)}&=&-2\sumintf_{\{P\}}\ln P^2 \nn
&=&
-\frac{7\pi^2}{180}T^4\[1+\frac{120}{7}\hmu^2+\frac{240}{7}\hmu^4\]
\;.
\ee
This is the free energy of an ideal gas consisting of a single massless fermion.

\subsubsection*{Soft contributions}

The soft contributions in the diagrams ${\cal F}_{1a+1b}^g$, ${\cal F}_{2d}^g$, and ${\cal F}_{3m}^g$ arise from the $P_0=0$ term in the sum-integral.  At soft momentum $P=(0,{\bf p})$, the HTL self-energy functions reduce to $\Pi_T(P) = 0$ and $\Pi_L(P) = m_D^2$.  The transverse term vanishes in dimensional regularization because there is no momentum scale in the integral over ${\bf p}$.  Thus the soft contributions come from the longitudinal term only and read
\be
{\cal F}^{g(s)}_{\rm 1a+1b}
&=&\frac{1}{2}T\int\limits_p\ln\left(p^2+m_D^2\right) = 
- \frac{m_D^3T}{12\pi}
\left( \frac{\Lambda_g}{2 m_D} \right)^{2 \epsilon}\left[
1+\frac{8}{3}\epsilon 
\right] , 
\label{f1as}
\ee
\be
{\cal F}^{g(s)}_{\rm 2d}&=&
-\frac{1}{2}m_D^2T\int\limits_p\frac{1}{p^2+m_D^2} =
\frac{m^3_DT}{8\pi} \left( \frac{\Lambda_g}{2 m_D} \right)^{2 \epsilon}
\left[1 + 2 \epsilon  
 \right]
\label{f2ds}
 ,
\ee
\be
{\cal F}^{g(s)}_{\rm 3m}&=& - \frac{1}{4}m_D^4T\int\limits_p\frac{1}{(p^2+m_D^2)^2} 
= - \frac{m^3_DT}{32\pi}
\label{f3ms}
\;.
\ee
The total soft contribution from eqs.~(\ref{f1as})-(\ref{f3ms}) is
\be
{\cal F}_{\rm 1a+1b+2d+3m}^{g(s)}&=&-\frac{m_D^3T}{96\pi}\left( \frac{\Lambda_g}{2 m_D} \right)^{2 \epsilon}\left[
1+\frac{8}{3}\epsilon
\right] .
\ee 
There are no soft contributions from the leading-order fermion diagrams or HTL counterterms (polarization and self-energy insertions).

\subsection{Two-loop sum-integrals}

For hard momenta, the self-energies are suppressed by $m_D/T$ and $m_q/T$ relative to the inverse free propagators, so we can expand in powers of $\Pi_T$, $\Pi_L$, and $\Sigma$.  As was the case for the one-loop contributions, we once again treat the polarization and self-energy insertion NNLO diagrams as two-loop graphs in order to simplify the resulting expressions.

\subsubsection*{(hh) contribution}

We first consider the contribution from fermionic diagrams.  The $(hh)$ contribution from ${\cal F}_{\rm 2a}^f$ and ${\cal F}_{\rm 2b}^f$ reads

\be
{\cal F}_{\rm 2a+2b}^{f(hh)}&=&(d-1)g^2\left[\ \sumintff_{\{PQ\}}\frac{1}{P^2Q^2}
-\sumintbf_{P\{Q\}}\frac{2}{P^2Q^2}\right] 
\nonumber \\ &&
+ 2m_D^2g^2\sumintbf_{P\{Q\}}\left[\frac{1}{p^2P^2Q^2}
{\cal T}_P+\frac{1}{P^4Q^2}
- \frac{d-2}{d-1}\frac{1}{p^2P^2Q^2}
\right]
\nonumber \\ &&
+ m_D^2g^2\sumintff_{\{PQ\}}
\left[ \frac{d+1}{d-1}\frac{1}{P^2Q^2r^2}
-\frac{4d}{d-1}\frac{q^2}{P^2Q^2r^4}-\frac{2d}{d-1}
\frac{P\!\cdot\!Q}{P^2Q^2r^4}\right]{\cal T}_R  
\nonumber \\ &&
+  
m_D^2g^2\sumintff_{\{PQ\}}\left[ \frac{3-d}{d-1}\frac{1}{P^2Q^2R^2}+
\frac{2d}{d-1}\frac{P\!\cdot\! Q}{P^2Q^2r^4}
-\frac{d+2}{d-1}
\frac{1}{P^2Q^2r^2} 
+\frac{4d}{d-1}\frac{q^2}{P^2Q^2r^4}\right.
\nonumber \\ &&
- \left.\frac{4}{d-1}\frac{q^2}{P^2Q^2r^2R^2} 
\right] +2m_q^2g^2(d-1)\sumintff_{\{PQ\}}\left[ \frac{d+3}{d-1}\frac{1}{P^2Q^2R^2}
- \frac{2}{P^2Q^4} 
+\frac{r^2-p^2}{q^2P^2Q^2R^2}\right]
\nonumber \\ &&
+ 2m_q^2g^2(d-1)\sumintff_{\{PQ\}}\left[ \frac{1}{P^2Q_0^2Q^2}
+\frac{p^2-r^2}{P^2q^2Q_0^2R^2}
\right] {\cal T}_Q
\nonumber \\ &&
+ 2m_q^2g^2(d-1)\sumintbf_{P\{Q\}} \left[\frac{2}{P^2Q^4}
-\frac{1}{P^2Q_0^2Q^2}{\cal T}_Q\right] .
\label{f2ab}
\ee
We consider next the $(hh)$ contributions from ${\cal F}_{\rm 3d}^f$ and ${\cal F}_{\rm 3f}^f$.
The easiest way to calculate these term, is to expand the two-loop
diagrams ${\cal F}_{\rm 2a}^f$ and ${\cal F}_{\rm 2b}^f$ to first order in $m_D^2$. This yields
\be
{\cal F}_{\rm 3d+3f}^{f(hh)}&=&
- 2m_D^2g^2\sumintbf_{P\{Q\}}
\left[\frac{1}{p^2P^2Q^2}{\cal T}_P+\frac{1}{P^4Q^2}
- \frac{d-2}{d-1}\frac{1}{p^2P^2Q^2}
\right] 
\nonumber \\ &&
- m_D^2g^2\sumintff_{\{PQ\}}
\left[ \frac{d+1}{d-1}\frac{1}{P^2Q^2r^2}
-\frac{4d}{d-1}\frac{q^2}{P^2Q^2r^4}-\frac{2d}{d-1}
\frac{P\!\cdot\!Q}{P^2Q^2r^4}\right]{\cal T}_R  
\nonumber \\ &&
- m_D^2g^2\sumintff_{\{PQ\}}\left[ \frac{3-d}{d-1}\frac{1}{P^2Q^2R^2}
+ \frac{2d}{d-1}\frac{P\!\cdot\! Q}{P^2Q^2r^4}
-\frac{d+2}{d-1}
\frac{1}{P^2Q^2r^2} \right.
\nonumber \\ && \left.
\hspace{4cm}
+\frac{4d}{d-1}\frac{q^2}{P^2Q^2r^4}
-\frac{4}{d-1}\frac{q^2}{P^2Q^2r^2R^2} 
\right] .
\label{f3df}
\ee
Next we consider the $(hh)$ contribution from the diagrams ${\cal F}_{3e}^f,{\cal F}_{3g}^f, {\cal F}_{3k}^f$ and ${\cal F}_{3l}^f$
\be
{\cal F}_{\rm 3e+3g+3k+3l}^{f(hh)}&=&
- 2m_q^2g^2(d-1)\left( \;\; \sumintff_{\{PQ\}}\left[ \frac{1}{P^2Q_0^2Q^2}
+\frac{p^2-r^2}{P^2q^2Q_0^2R^2}
\right] {\cal T}_Q \right.
\nonumber \\ && \left.
+ \; \sumintbf_{P\{Q\}} \left[\frac{2}{P^2Q^4}
+\frac{1}{P^2Q_0^2Q^2}{\cal T}_Q\right]\right. 
\nn
&& 
\left. + \sumintff_{\{PQ\}}\left[ \frac{d+3}{d-1}\frac{1}{P^2Q^2R^2}
- \frac{2}{P^2Q^4} 
+\frac{r^2-p^2}{q^2P^2Q^2R^2}\right] \right).
\label{f3egkl}
\ee
The sum of eqs.(\ref{f2ab})-(\ref{f3egkl}) is
\be
{\cal F}_{\rm 2a+2b+3d+3e+3f+3g+3k+3l}^{f(hh)}&=&(d-1)g^2\left[\ \sumintff_{\{PQ\}}\frac{1}{P^2Q^2}
-\sumintbf_{P\{Q\}}\frac{2}{P^2Q^2}\right] 
\nn
&=&\frac{\pi^2}{72}\frac{\alpha_s}{\pi}T^4\lb1+12\hmu^2\rb\lb5+12\hmu^2\rb \, .
\ee
For completeness, the hard-hard contribution coming from two-loop pure-glue diagrams is \cite{3loopglue2}
\be
{\cal F}_{\rm 2a+2b+2c+3h+3i+3j+3k+3l}^{g(hh)}&=&\frac{1}{4}(d-1)^2g^2 \sumintbb_{PQ}\frac{1}{P^2Q^2}\nn
&=&\frac{\pi^2}{36}\frac{\alpha_s}{\pi}T^4
\ee

\subsubsection*{(hs) contribution}

In the $(hs)$ region, one gluon momentum is soft but the fermionic momentum is always hard.  The terms that contribute through order $g^2 m_D^3 T$ and $g^2m_q^2m_DT$ from ${\cal F}_{2a}^f$ and ${\cal F}_{2b}^f$ were calculated in ref.~\cite{andersen4,andersen5,najmul2} and read
\be
{\cal F}_{2a+2b}^{f(hs)}&=&2g^2T\int\limits_p\frac{1}{p^2+m^2_D}
\sumintf_{\{Q\}}\left[
\frac{1}{Q^2}-\frac{2q^2}{Q^4}\right]
\nn
&& + 2m_D^2g^2T\int\limits_p\frac{1}{p^2+m_D^2}
\sumintf_{\{Q\}}
\left[\frac{1}{Q^4}
-\frac{2}{d}(3+d)\frac{q^2}{Q^6}+\frac{8}{d}\frac{q^4}{Q^8}
\right]
\nn 
&&
- 4m_q^2g^2T\int\limits_p\frac{1}{p^2+m_D^2}
\sumintf_{\{Q\}}\left[\frac{3}{Q^4}
-\frac{4q^2}{Q^6} -\frac{4}{Q^4} {\cal T}_Q
-\frac{2}{Q^2}\bigg\langle \frac{1}{(Q\!\cdot\!Y)^2} \bigg\rangle_{\!\!\bf \hat y}
\right].
\label{2a2bhs}
\ee
The $(hs)$ contribution from diagrams ${\cal F}_{\rm 3d}^f$ and ${\cal F}_{\rm 3f}^f$ can again be calculated from the diagrams ${\cal F}_{2a}^f$ and ${\cal F}_{2b}^f$ by Taylor expanding their contribution to first order in $m_D^2$. This yields
\be
{\cal F}_{\rm 3d+3f}^{f(hs)}&=&
2m_D^2g^2T\int\limits_p\frac{1}{(p^2+m_D^2)^2}
\sumintf_{\{Q\}}\left[\frac{1}{Q^2}-\frac{2q^2}{Q^4}\right]
\nn
&&-2m_D^2g^2T\int\limits_p\frac{p^2}{(p^2+m^2_D)^2}\sumintf_{\{Q\}}
\left[\frac{1}{Q^4}-\frac{2}{d}(3+d)\frac{q^2}{Q^6}+\frac{8}{d}\frac{q^4}{Q^8}
\right]
\nn && 
-4m_D^2m_q^2g^2T\int\limits_p\frac{1}{(p^2+m_D^2)^2}\sumintf_{\{Q\}}
\left[\frac{3}{Q^4}
-\frac{4q^2}{Q^6}
-\frac{4}{Q^4}{\cal T}_{Q}-\frac{2}{Q^2}
\bigg\langle
 \frac{1}{(Q\!\cdot\!Y)^2} \bigg\rangle_{\!\!\bf \hat y}
\right] \! . \hspace{9mm}
\label{3d3fhs}
\ee
We also need the $(hs)$ contributions from the diagrams ${\cal F}_{3e}^f,{\cal F}_{3g}^f, {\cal F}_{3k}^f$ and ${\cal F}_{3l}^f$. Again we calculate these contributions by expanding the two-loop diagrams ${\cal F}_{2a}^f$ and ${\cal F}_{2b}^f$ to first order in $m_q^2$. This yields
\be
{\cal F}_{3e+3g+3k+3l}^{f(hs)}&=&4m_q^2g^2T\int_{p}\frac{1}{p^2+m_D^2}
\sumintf_{\{Q\}}\left[\frac{3}{Q^4}
-\frac{4q^2}{Q^6} - \frac{4}{Q^4} {\cal T}_Q
-\frac{2}{Q^2}\bigg\langle \frac{1}{(Q\!\cdot\!Y)^2} \bigg\rangle_{\!\!\bf \hat y}
\right] \! . \hspace{9mm}
\label{3egklhs}
\ee
The sum of eqs.~(\ref{2a2bhs})-(\ref{3egklhs}) is
\be
\hspace{6mm} &\;& \hspace{-1cm} {\cal F}_{2a+2b+3d+3e+3f+3g+3k+3l}^{f(hs)}
\nn
&=&2g^2 T\left[\int_p\frac{1}{p^2+m_D^2}+m_D^2\int_p\frac{1}{\lb p^2+m_D^2\rb^2}
\right]\sumintf_{\{Q\}}\left[\frac{1}{Q^2}-\frac{2q^2}{Q^4}\right]\nn
&+&2g^2m_D^2T \[\int_p\frac{1}{p^2+m_D^2}-\int_p\frac{p^2}{\lb p^2+m_D^2\rb^2}\]
\sumintf_{\{Q\}}
\left[\frac{1}{Q^4}-\frac{2}{d}(3+d)\frac{q^2}{Q^6}+\frac{8}{d}\frac{q^4}{Q^8}
\right]
\nn
&-& 4m_q^2m_D^2T\int_p\frac{1}{\lb p^2+m_D^2\rb^2}
\sumintf_{\{Q\}}\left[\frac{3}{Q^4}
-\frac{4q^2}{Q^6} -\frac{4}{Q^4} {\cal T}_Q
-\frac{2}{Q^2}\bigg\langle \frac{1}{(Q\!\cdot\!Y)^2} \bigg\rangle_{\!\!\bf \hat y}
\right]
\nn
&=&-2(d-1)g^2 T\left[\int_p\frac{1}{p^2+m_D^2}+m_D^2\int_p\frac{1}{\lb p^2+m_D^2\rb^2}
\right]\sumintf_{\{Q\}}\frac{1}{Q^2}
\nn
&-&\frac{d-1}{3}g^2m_D^4T\int_p\frac{1}{\lb p^2+m_D^2\rb^2}\sumintf_{\{Q\}}\frac{1}{Q^4}-8\frac{d-3}{d-1}
g^2m_q^2m_D^2T\int_p\frac{1}{\lb p^2+m_D^2\rb^2}\sumintf_{\{Q\}}\frac{1}{Q^4}\ ,
\nn
&=&-\frac{1}{12}\alpha_s\lb1+12\hmu^2\rb m_DT^3
-\frac{\alpha_s}{4\pi^2}m_q^2m_D^2T
\nonumber \\
&& \hspace{2cm}
-\frac{\alpha_s}{48\pi^2}\[\frac{1}{\epsilon}-1-\aleph(z)\]
\left(\frac{\Lambda}{4\pi T}\right)^{2\epsilon}\left(\frac{\Lambda}{2m_D}\right)^{2\epsilon}\, .
\ee

\subsection{Three-loop sum-integrals}

We now list the mass-expanded sum-integrals necessary at three loops.  As before we organize the
contributions according to whether the momentum flowing in a given propagator is hard or soft.

\subsubsection*{(hhh) contribution}

The $(hhh)$ contributions from diagrams ${\cal F}_{3a}^{f}$ and ${\cal F}_{3b}^{f}$ are
\begin{eqnarray}
 {\cal F}_{3a}^{f(hhh)}&=&g^4(d-1)\Biggl[4\Big(\sum\bsa\int\limits_{P}\ \frac{1}{P^2}
   -2\sum\bsa\!\int\limits_{\{P\}}\frac{1}{P^2}\Big)\sum\bsa\!\!\!\int\limits_{\{QR\}}\frac{1}{Q^2R^2(Q+R)^2}
   \nonumber \\
&&\hspace{-1.5cm}
-\frac{1}{2}(d-7)\sum\bs\int\limits_{\{PQR\}}\frac{1}{P^2Q^2R^2(P+Q+R)^2}
\nonumber\\
&&\hspace{-1.5cm}
+(d-3)\sum\bs\int\limits_{\{PQR\}}\frac{1}{P^2Q^2(P-R)^2
(Q-R)^2}  + 2 \sum\bs\int\limits_{\{PQ\}R}\frac{(P-Q)^2}{P^2Q^2R^2(P-R)^2(Q-R)^2}\Bigg] \nn
&&\hspace{-1.5cm} = g^4(d-1)\Big\{4\lb{\cal I}_1^0-2\tilde{\cal I}_1^0\rb\tilde\tau -\frac{1}{2}\lb d-7\rb{\cal N}_{0,0}
 +(d-3)\widetilde M_{0,0} + 2{\cal N}_{1,-1}\Big\}.
\end{eqnarray}
\begin{eqnarray}
 {\cal F}_{3b}^{f(hhh)}&=&-g^4(d-1)^2\Bigg[\Big(\sum\bsa\int\limits_{P}\ \frac{1}{P^2}
   -\sum\bsa\!\int\limits_{\{P\}}\frac{1}{P^2}\Big)^2\sum\bsa\!\int\limits_{\{P\}}\frac{1}{P^4}-2\sum\bs\int\limits_{\{PQR\}}
    \frac{1}{P^2Q^2R^2(Q+R)^2}   \nonumber\\
&& +\sum\bs\int\limits_{\{PQR\}}\frac{1}{P^2Q^2(P-R)^2(Q-R)^2}+ \sum\bs\int\limits_{\{PQR\}}\frac{(P-Q)^2}{P^2Q^2R^2(P-R)^2(Q-R)^2}
\Biggr] \nn
&=&-g^4(d-1)^2 \Big\{\lb{\cal I}_1^0-\tilde{\cal I}_1^0\rb^2\tilde{\cal I}_2^0-2\tilde{\cal I}_1^0\tilde{\tau}
+\widetilde{\cal M}_{0,0}+\widetilde{\cal M}_{1,-1}\Big\}.
\end{eqnarray}
Expressions for ${\cal F}_{3a}^f$, ${\cal F}_{3b}^f$, and ${\cal F}_{3c}^f$ can be found in ref.~\cite{vuorinen1,vuorinen2}.  The results are
\begin{eqnarray}
{\cal F}_{3a}^{f(hhh)}+{\cal F}_{3b}^{f(hhh)} && \nonumber \\
&& \hspace{-2cm} = 
\frac{\alpha_s^2T^4}{ 192}\Bigg[35-32\frac{\zeta'(-1)}{\zeta(-1)}+472 \hat\mu^2
+384  \frac{\zeta'(-1)}{\zeta(-1)} \hat\mu ^2+1328  \hat\mu^4
    \nonumber\\
&& \hspace{-1.5cm}
+64\Big(-36i\hat\mu\aleph(2,z)+6(1+8\hat\mu^2)\aleph(1,z)+3i\hat\mu(1+4\hat\mu^2)\aleph(0,z)\Big)\Bigg] .
\end{eqnarray}
\begin{eqnarray}
 {\cal F}_{3c}^{f(hhh)} &=& -\frac{ 5\alpha_s^2 }{216}T^4\left(\frac{\Lambda}{4\pi T}\right)^{6\epsilon}\Bigg[ \left(1 +
\frac{72}{5}\hat\mu^2+\frac{144}{5}\hat\mu^4\right)\frac{1}{\epsilon}
+ \frac{31}{10}+\frac{6}{5}\gamma_E - \frac{68}{25}\frac{\zeta'(-3)}{\zeta(-3)}\nn
&+& \frac{12}{5}(25+12\gamma_E)\hat\mu^2 + 120\hat\mu^4
 - \frac{8}{5}(1+12\hat\mu^2)\frac{\zeta'(-1)}{\zeta(-1)} 
\nn
 &-&  \frac{144}{5}\Big[8\aleph(3,z)+3\aleph(3,2z) +12 i \hat\mu\,(\aleph(2,z)+\aleph(2,2z)) - (3+20\hat\mu^2)\aleph(1,z)
\nn
 &-&
i \hat\mu(1+12\hat\mu^2)\,\aleph(0,z)  
 -12\hat\mu^2\aleph(1,2z) \Big]\Bigg] \, ,
\end{eqnarray}
where $\aleph(n,z)$ is defined in appendix \ref{app:aleph}.  The $(hhh)$ contribution proportional to $c_A s_F$ is 
\begin{eqnarray}
&-&\frac{1}{2}{\cal F}_{3a}^{f(hhh)}+{\cal F}_{3m+3n}^{f(hhh)}\nonumber\\
&=&          g^2(d-1)\Bigg\{2(d-5)\sumintbbf_{PQ\{R\}}\frac{1}{P^4Q^2R^2}+\frac{1}{2}(d-3)
             \sumintfff_{\{PQR\}}\frac{1}{P^2Q^2(P-Q)^2(Q-R)^2}\nonumber\\
&&\qquad\qquad  -\frac{1}{4}(d-7)\sumintbff_{\{PQ\}R}\frac{1}{P^2Q^2(P-R)^2(Q-R)^2}+\sumintbff_{\{PQ\}R}
\frac{(P-Q)^2}{P^2Q^2R^2(P-R)^2(Q-R)^2}\nn
&&\qquad\qquad -(d-3)\sumintbff_{P\{QR\}}\frac{1}{P^2Q^2R^2(Q-R)^2}-2\sumintfff_{\{PQR\}}\frac{1}{P^2Q^2R^2(Q-R)^2}
\nonumber\\ 
&&\qquad\qquad +2\sumintfff_{\{PQR\}}
      \frac{R^4}{P^2Q^2(P-Q)^4(Q-R)^4}\Bigg\} 
\nn
&=& g^2(d-1)\Bigg[2(d-5){\cal I}_2^0{\cal I}_1^0\tilde {\cal I}_1^0+\frac{1}{2}(d-3)\widetilde {\cal M}_{00}-\frac{1}{4}
(d-7){\cal N}_{00}+{\cal N}_{1,-1}-(d-3){\cal I}_1^0\tilde\tau-2\tilde {\cal I}_1^0\tilde\tau\Bigg]\nn
&=&         -\frac{25\alpha_s^2T^4}{864}\Bigg[\left(1+\frac{72}{25}\hat\mu^2-\frac{1584}{25}
             \hat\mu^4\right)\lb\frac{1}{\epsilon}
           +6\ln\frac{\hat\Lambda}{2}\rb-\frac{369}{250}\left(1+\frac{2840}{123}\hat\mu^2+\frac{28720}{123}\hat\mu^4\right)
            \nonumber\\
&&\qquad\qquad    +\frac{48}{25}\left(1+12\hat\mu^2\right)\gamma_E+\frac{536}{125}\frac{\zeta'(-3)}{\zeta(-3)}
          +\frac{32}{25}\lb1+6\hat\mu^2\rb\frac{\zeta'(-1)}{\zeta(-1)}+\frac{288}{25}\Big[26\aleph(3,z)
\nonumber\\           
&&\qquad\qquad    +\left(3-68\hat\mu^2\right)\aleph(1,z)+72i\hmu\aleph(2,z)+2i\hmu\aleph(0,z)\Big]\Bigg] ,          
\end{eqnarray}
where the integrals appearing above are evaluated in app.~\ref{app:threeloopsumints}.
Finally, we note that there is no $(hhh)$ contribution from ${\cal F}_{3o}^f$ since this is a purely HTL diagram.

\subsubsection*{(hhs) contribution}

The $(hhs)$ contribution to the ${\cal F}_{\rm 3a}$, ${\cal F}_{\rm 3b}$, and ${\cal F}_{\rm 3c+3j}$ are
\begin{eqnarray}
 {\cal F}_{\rm 3a}^{f(hhs)}&=&
2(d-1)g^4T\int\limits_p\frac{1}{p^2+m_D^2}\Bigg[\ \ 
\sumintbf_{Q\{R\}}\frac{4Q_0R_0}{Q^2R^4(Q+R)^2}
\nonumber \\
&& \hspace{4cm}
-\sumintff_{\{QR\}}\frac{1}{Q^2R^2(Q-R)^2}
+\sumintff_{\{QR\}}\frac{2Q_0R_0}{Q^4R^4}\Bigg] , 
\label{3ahhs} 
\end{eqnarray}
\begin{eqnarray}
 {\cal F}_{\rm 3b}^{f(hhs)}&=&
2(d-1)g^4T\int\limits_p\frac{1}{p^2+m_D^2}\left[\ \ 
\sumintbf_{Q\{R\}}-\frac{4Q_0R_0}{Q^2R^4(Q+R)^2}+\sumintff_{\{QR\}}\frac{1}{Q^2R^2(Q-R)^2}\right.
\nonumber\\
&&
\hspace{2cm}
+ \left.(d-3)\sumintf_{\{Q\}}\frac{1}{Q^4}
\left(\sumintb_{R}\frac{1}{R^2}-\sumintf_{\{R\}}\frac{1}{R^2}\right)\right] , 
\end{eqnarray}
\begin{eqnarray}
{\cal F}_{3c+3j}^{f(hhs)}&=&
-4g^4T\int\limits_p\frac{1}{(p^2+m_D^2)^2}
\left[\ \sumintf_{\{Q\}}\left(
\frac{1}{Q^2}-\frac{2q^2}{Q^4}\right)
\right]^2
\nn && 
\hspace{-1cm}
+8g^4T\int\limits_p\frac{p^2}{(p^2+m_D^2)^2}
\sumintf_{\{Q\}}\left[\frac{1}{Q^2}-\frac{2q^2}{Q^4}\right]
\sumintf_{\{R\}}\left[
\frac{1}{R^4}-\frac{2}{d}(3+d)
\frac{r^2}{R^6}+\frac{8}{d}\frac{r^4}{R^8}
\right]\nn
&&
\hspace{-1cm}
-16m_q^2g^4T\int\limits_{p}\frac{1}{(p^2+m_D^2)^2}
\sumintf_{\{Q\}}\left[\frac{1}{Q^2}-\frac{2q^2}{Q^4}\right]
\sumintf_{\{R\}}\left[
\frac{3}{R^4}-\frac{4r^2}{R^6}
-\frac{4}{R^4}{\cal T}_R-\frac{2}{R^2}
\bigg\langle\frac{1}{(R\!\cdot\!Y)^2} \bigg\rangle_{\!\!\bf \hat y}
\right]\;
\nn
&& \hspace{-1cm}
 = -4g^4T(d-1)^2\int\limits_p\frac{1}{(p^2+m_D^2)^2}
\sumintff_{\{QR\}}
\frac{1}{Q^2R^2}
+\frac{4}{3}g^4T(d-1)^2\int\limits_p\frac{p^2}{(p^2+m_D^2)^2}
\sumintff_{\{QR\}}\frac{1}{Q^2R^4}
\nn
&&+32m_q^2g^4T(d-3)
\int\limits_{p}\frac{1}{(p^2+m_D^2)^2}
\sumintff_{\{QR\}}\frac{1}{Q^2R^4}\;.
\end{eqnarray}
Computing the necessary sum-integrals one finds
\begin{eqnarray}
 {\cal F}^{f(hhs)}_{3a+3b}=\frac{\alpha_s^2 m_DT^3}{4\pi}(1+12\hat\mu^2) \, ,
\end{eqnarray}
and 
\begin{eqnarray}
 {\cal F}^{f(hhs)}_{3c+3j}&=&
\frac{\alpha_s^2m_DT^3}{12\pi}\left[\frac{1+12\hat\mu^2}{\epsilon}+(1+12\hat\mu^2)\left(\frac{4}{3}
-\aleph(z)\right)+24\aleph(1,z)\right]
\nonumber\\
&-&\frac{\pi\alpha_s^2T^5}{18m_D}\left(1+12\hat\mu^2\right)^2-\frac{\alpha_s^2m_q^2T^3}{3\pi m_D}\left(1+12\hat\mu^2\right) .
\end{eqnarray}
Similarly, one obtains
\begin{eqnarray}
&& -\frac{1}{2}{\cal F}_{3a}^{f(hhs)}+{\cal F}_{3m+3n+3o}^{f(hhs)} 
\nonumber \\ && \hspace{.5cm} 
= g^2T(d-1)\Bigg\{2(d-1)^2\int\limits_p\frac{1}{(p^2+m_D^2)^2}\sumintbf_{Q\{R\}}\frac{1}{Q^2R^2}
\nonumber \\ && \hspace{1cm} 
+\frac{1}{2}(d-3)\int\limits_p\frac{1}{(p^2+m_D^2)}\sumintff_{\{QR\}}\frac{1}{Q^2R^2(Q-R)^2}
\nonumber\\
&&\ \hspace{1cm} - \int\limits_p\frac{1}{(p^2+m_D^2)}\sumintff_{\{QR\}}\frac{2Q_0R_0}{Q^4R^4}  
-\frac{1}{3}\left(d^2-11d+46\right)\int\limits_p\frac{p^2}{(p^2+m_D^2)^2}\sumintbf_{Q\{R\}}\frac{1}{Q^4R^2}
\nn                     
&& \hspace{1cm} -\frac{1}{3}(d-1)^2\int\limits_p\frac{p^2}{(p^2+m_D^2)^2}\sumintbf_{Q\{R\}}\frac{1}{Q^2R^4}
\nonumber\\
&& \hspace{1cm}  +4m_q^2(d-1)\int\limits_p\frac{1}{(p^2+m_D^2)^2}\sumintb_{Q}\frac{1}{Q^2}\sumintf_{\{R\}}\left[\frac{3}{R^4}
                     -\frac{4r^2}{R^6}-\frac{4}{R^4}{\cal T}_R-\frac{2}{R^2}\left\langle\frac{1}{(R.Y)^2}\right\rangle_
                     {\hat{\bf y}}\right]\Bigg\} \, ,
\nonumber\\ && \hspace{.5cm} 
= g^2T(d-1)\Bigg\{2(d-1)^2\int\limits_p\frac{1}{(p^2+m_D^2)^2}\sumintbf_{Q\{R\}}\frac{1}{Q^2R^2}
\nonumber\\ && \hspace{1cm}
-\frac{1}{3}\left(d^2-11d+46\right)\int\limits_p\frac{p^2}{(p^2+m_D^2)}\sumintbf_{Q\{R\}}\frac{1}{Q^4R^2}
\nonumber\\ && \hspace{1cm}  
                     -\frac{1}{3}(d-1)^2\int\limits_p\frac{p^2}{(p^2+m_D^2)^2}\sumintbf_{Q\{R\}}\frac{1}{Q^2R^4}
          +  8m_q^2(d-3)\int\limits_p\frac{1}{(p^2+m_D^2)^2}\sumintbf_{Q\{R\}}\frac{1}{Q^2R^4}
                     \Bigg\}
\nonumber\\ && \hspace{.5cm}
= -\frac{\alpha_s m_DT^3}{48\pi}\left(\frac{\Lambda}{2m_D}\right)^{2\epsilon}\left(\frac{\Lambda}{4\pi T}\right)^{4\epsilon}
         \Bigg[\frac{1}{\epsilon} \left(7+132\hat\mu^2\right)+\frac{88}{3}+440\hat\mu^2+22\left(1+12\hat\mu^2\right)\gamma_E
\nonumber\\   
&&\hspace{1.5cm} -8\frac{\zeta'(-1)}{\zeta(-1)} +\ 4\aleph(z)+264\aleph(1,z)\Bigg]
-\frac{\pi\alpha_s^2T^5}{9m_D}\left(1+12\hat\mu^2\right)-\frac{\alpha_s^2}{3\pi m_D}m_q^2T^3 \, .
\end{eqnarray}

\subsubsection*{(hss) contribution}

The only three-loop diagram involving a fermionic line that has a $(hss)$ contribution is 
${\cal F}_{3n}^f$ which can be written as
\begin{eqnarray}
 {\cal F}_{3n}^{f(hss)}&=&g^4T^2\int\limits_{pq}\left[\frac{4}{\left(p^2+m_D^2\right)
       \left(q^2+m_D^2\right)\lb{\bf p}+{\bf q}\rb^2}-\frac{2}
      {\left(p^2+m_D^2\right) \left(q^2+m_D^2\right)^2}\right.
\nonumber\\
&&
\left.\hspace{1cm}-\frac{8 m_D^2}
      {\lb{\bf p}+{\bf q}\rb^2\left(p^2+m_D^2\right)
       \left(q^2+m_D^2\right)^2}\right]\sumintf_{\{R\}}\lb\frac{1}{R^2}-\frac{2r^2}{R^4}\rb
\nonumber\\
&=&-g^4T^2(d-1)\int\limits_{pq}\left[\frac{4}{\left(p^2+m_D^2\right)
       \left(q^2+m_D^2\right)\lb{\bf p}+{\bf q}\rb^2}-\frac{2}
      {\left(p^2+m_D^2\right) \left(q^2+m_D^2\right)^2}\right.
\nonumber\\
&&
\left.\hspace{5cm}-\frac{8 m_D^2}
      {\lb{\bf p}+{\bf q}\rb^2\left(p^2+m_D^2\right)
       \left(q^2+m_D^2\right)^2}\right]\sumintf_{\{R\}}\frac{1}{R^2}
\nonumber\\
&=&  \frac{\alpha_s^2T^4}{12}\left[\frac{1+12\hat\mu^2}{\epsilon}+2\left(1+12\hat\mu^2+12\aleph(1,z)\right)\right]
      \left(\frac{\Lambda}{2m_D}\right)^{4\epsilon}\left(\frac{\Lambda}{4\pi T}\right)^{2\epsilon}
\end{eqnarray}

\section{Sum-Integrals}
\label{app:sum-integrals}

We can define a set of ``master" sum-integrals as in~\cite{vuorinen1,vuorinen2}
\be
{\cal I}_n^m &=& \sumintb_{P}\frac{P_0^m}{P^{2n}}\ ,\\
{\widetilde{\cal I}}_n^m &=& \sumintf_{\{P\}}\frac{P_0^m}{P^{2n}}\ ,\\
\tilde{\tau}&=&\sumintff_{\{PQ\}}\frac{1}{P^2Q^2(P+Q)^2}\ ,\\
{\cal M}_{m,n}&=&\sumintbbb_{PQR}\frac{1}{P^2Q^2(R^2)^m[(P-Q)^2]^n(P-R)^2(Q-R)^2}\ ,\\
\widetilde{\cal M}_{m,n}&=&\sumintfff_{\{PQR\}}\frac{1}{P^2Q^2(R^2)^m[(P-Q)^2]^n(P-R)^2(Q-R)^2}\ ,\\
 {\cal N}_{m,n}&=&\sumintbff_{\{PQ\}R}\frac{1}{P^2Q^2(R^2)^m[(P-Q)^2]^n(P-R)^2(Q-R)^2} \ .
\ee

\subsection{One-loop sum-integrals}

The specific bosonic sun-integrals needed are
\begin{eqnarray}
{\cal I}_1^0 = \sumintb_{P}\ \frac{1}{P^2}=\frac{T^2}{12}\left(\frac{\Lambda_g}{4\pi T}\right)^{2\epsilon}\left[1+2\epsilon\left(1+\frac{\zeta'(-1)}
                      {\zeta(-1)}\right)\right] ,
\end{eqnarray}
\begin{eqnarray}
{\cal I}_2^0 = \sumintb_{P}\ \frac{1}{P^4}=\frac{1}{\left(4\pi\right)^2}\left(\frac{\Lambda_g}{4\pi T}\right)^{2\epsilon}\left[\frac{1}
{\epsilon}+2\gamma_E \right] .
\end{eqnarray}

The specific fermionic sun-integrals needed are
\begin{eqnarray}
\widetilde{\cal I}_1^0=\sumintf_{\{P\}}\frac{1}{P^2}=-\frac{T^2}{24}\left(\frac{\Lambda}{4\pi T}\right)^{2\epsilon}\left[1+12\hat\mu^2+2\epsilon
\left(1+12\hat\mu^2
       +12\aleph(1,z)\right)\right] ,
\end{eqnarray}
and
\begin{eqnarray}
\widetilde{\cal I}_2^0=\sumintf_{\{ P\} }\frac{1}{P^4}=\frac{1}{\left(4\pi\right)^2}\left(\frac{\Lambda}{4\pi T}\right)^{2\epsilon}\left[\frac{1}
{\epsilon}-\aleph(z)\right] .
\end{eqnarray}

Using the two basic one-loop sum-integrals above, we can construct other one-loop sum-integrals that will be necessary here as follows:
\begin{eqnarray}
 \sumintf_{\{P\}}\frac{1}{P^2}=\frac{2}{d}\sumintf_{\{P\}}\frac{p^2}{P^4}
\end{eqnarray}
\begin{eqnarray}
 \sumintf_{\{P\}}\frac{1}{P^4} &=& \frac{4}{d}\sumintf_{\{P\}}\frac{p^2}{P^6}=\frac{24}{d(d+2)}\sumintf_{\{P\}}\frac{p^4}{P^8} \, , \nn
&=& (d-1)\sumintf_{\{P\}}\frac{1}{P^4}{\cal T}_P=-\frac{2}{d-1}\sumintf_{\{P\}}\frac{1}{P^2}\left\langle\frac{1}{(P\cdot Y)^2}\right\rangle_{\hat{\bf y}} \, .
\end{eqnarray}
This allows us to compute the following sum-integrals
\begin{eqnarray}
\sumintf_{\{P\}}\frac{p^2}{P^4}=-\frac{T^2}{16}\left(\frac{\Lambda}{4\pi T}\right)^{2\epsilon}\left[1+12\hat\mu^2
        +\frac{4}{3}\epsilon\left(1+12\hat\mu^2
       +18\aleph(1,z)\right)\right] ,
\end{eqnarray}
\begin{eqnarray}
\sumintf_{\{P\}}\frac{1}{P^4}{\cal T}_P=\frac{1}{2}\frac{1}{\left(4\pi\right)^2}\left(\frac{\Lambda}{4\pi T}\right)^{2\epsilon}
\left[\frac{1}{\epsilon}+1-\aleph(z)\right] ,
\end{eqnarray}
\begin{eqnarray}
\sumintf_{\{P\}}\frac{p^2}{P^6}=\frac{3}{4}\frac{1}{\left(4\pi\right)^2}\left(\frac{\Lambda}{4\pi T}\right)^{2\epsilon}
\left[\frac{1}{\epsilon}-\frac{2}{3}
            -\aleph(z)\right] ,
\end{eqnarray}
\begin{eqnarray}
\sumintf_{\{P\}}\frac{p^4}{P^8}=\frac{5}{8}\frac{1}{\left(4\pi\right)^2}\left(\frac{\Lambda}{4\pi T}\right)^{2\epsilon}
\left[\frac{1}{\epsilon}-\frac{16}{15}-\aleph(z)\right] ,
\end{eqnarray}
\begin{eqnarray}
\sumintf_{\{P\}}\frac{1}{P^2}\left\langle\frac{1}{(P.Y)^2}\right\rangle_{\hat{\bf y}}=-\frac{1}{\left(4\pi\right)^2}
      \left(\frac{\Lambda}{4\pi T}
      \right)^{2\epsilon}\left[\frac{1}{\epsilon}-1-\aleph(z)\right] ,
\end{eqnarray}
\begin{eqnarray}
 \sumintf_{\{P\}}\frac{P_0}{P^4}=\frac{1}{\left(4\pi\right)}
      \left(\frac{\Lambda}{4\pi T}\right)^{2\epsilon}\left[i\hat\mu+\ \aleph(0,z)\epsilon\right] .
\end{eqnarray}

\subsection{Two-loop sum-integrals}
\label{app:twoloopsumints}

For the purposes of this paper we only need one new two-loop sum-integral
\be
\tilde{\tau}=\sumintff_{\{PQ\}}\frac{1}{P^2Q^2(P+Q)^2}=-\frac{T^2}{\lb4\pi\rb^2}\lb\frac{\Lambda}{4\pi T}\rb^{4\epsilon}
\[\frac{\hmu^2}{\epsilon}+2\hmu^2-2i\hmu\aleph[0,z]\] \, .
\ee

\subsection{Three-loop sum-integrals}
\label{app:threeloopsumints}

%
%
The three-loop sum-integrals necessary are
\begin{eqnarray}
{\cal M}_{00}&=&\sumintbbb_{PQR}\frac{1}{P^2Q^2R^2\left(P+Q+R\right)^2}
\nn
&=&\frac{1}{(4\pi)^2}\left(\frac{T^2}{12}\right)^2\left(\frac{\Lambda_g}
         {4\pi T}\right)^{6\epsilon}\left[\frac{6}{\epsilon}+\frac{182}{5}-12\frac{\zeta'(-3)}{\zeta(-3)}+48\frac{\zeta'(-1)}
         {\zeta(-1)}\right] .
\end{eqnarray}
%
%
\begin{eqnarray}
{\cal N}_{00}&=&\sumintfff_{\{PQR\}}\frac{1}{P^2Q^2R^2\left(P+Q+R\right)^2}\nn
             &=&\frac{1}{(4\pi)^2}\left(\frac{T^2}{12}\right)^2\left(\frac{\Lambda}
         {4\pi T}\right)^{6\epsilon}\Bigg[\frac{3}{2\epsilon}\left(1+12\hat\mu^2\right)^2+\frac{173}
          {20}+210\hat\mu^2+1284\hat\mu^4-\frac{24}{5}\frac{\zeta'(-3)}{\zeta(-3)}\nn
&+& \! 144\Big[(1+8\hat\mu^2)\aleph(1,z)+4
          \hat\mu^2\aleph(1,2z)
      -4i\hat\mu\left[\aleph(2,z)+\aleph(2,2z)\right]-2\aleph(3,z)-\aleph(3,2z)
\Big]\Bigg] .
\nn
\end{eqnarray}
%
%
\begin{eqnarray}
 \widetilde{\cal M}_{00}&=&\sumintbbf_{PQ\{R\}}\frac{1}{P^2Q^2R^2\left(P+Q+R\right)^2}\nn
                        &=&-\frac{1}{(4\pi)^2}\left(\frac{T^2}{12}\right)^2\left(\frac{\Lambda}
         {4\pi T}\right)^{6\epsilon}\Bigg[\frac{3}{4\epsilon}\left(1+24\hat\mu^2-48\hat\mu^4\right)+\frac{179}
          {40}+111\hat\mu^2-210\hat\mu^4
\nn&+&
      48\frac{\zeta'(-1)}{\zeta(-1)}\hat\mu^2+\frac{24}{5}\frac{\zeta'(-3)}{\zeta(-3)}
         +72\Big[(1-8\hat\mu^2)\aleph(1,z)+6\aleph(3,z)+12i\hat\mu\aleph(2,z)\Big]\Bigg] .
         \nn
\end{eqnarray}
%
%
\begin{eqnarray}
{\cal N}_{1,-1} &=&
-\frac{1}{2\lb4\pi\rb^2}\lb\frac{T^2}{12}\rb^2\left(\frac{\Lambda}
         {4\pi T}\right)^{6\epsilon}\bigg[\frac{3}{2\epsilon}\lb1+12\hmu^2\rb\lb1-4\hmu^2\rb+
+\frac{173}{20} + 114\hmu^2\nn
&+&132\hmu^4
- \frac{12}{5}\Zc - 96\hmu^2\Za
- 144\Big[2\aleph(3,z) + 2\aleph(3,2z) - 4i \hmu\,\aleph(2,z) \nn
&+& 8i\hmu\,\aleph(2,2z)
- \lb1-4\hmu^2\rb\aleph(1,z)
- 8\hmu^2\aleph(1,2z) -\frac{1}{3}i\hmu\lb1+12\hmu^2\rb\aleph(0,z)\Big] \bigg] \, .
\nn 
\end{eqnarray}
%
%
\begin{eqnarray}
H_3&=& \sumintbbf_{\{P\}QR}\frac{Q\cdot R}{P^2Q^2R^2\left(P+Q\right)^2\left(P+R\right)^2}
\nn
&=&\frac{1}{(4\pi)^2}\left(\frac{T^2}{12}\right)^2
                \left(\frac{\Lambda}{4\pi T}\right)^{6\epsilon}\Bigg[\frac{3}{8\epsilon}\left(1+12\hat\mu^2\right)^2
                +\frac{361}{160}-\frac{3}{5}\frac{\zeta'(-3)}{\zeta(-3)}
\nonumber\\
&+&\frac{141}{4}\hat\mu^2+\frac{501}{2}\hat\mu^4
         -9\Bigg\{\left(\frac{1}{8}+\hat\mu^2+2\hat\mu^4\right)\aleph(z)+2i\hat\mu\left(1+4\hat\mu^2\right)\aleph(0,z) 
\nonumber\\
&+&2\left(1-12\hat\mu^2\right)\aleph(1,z)+24i\hat\mu\aleph(2,z)+16\aleph(3,z)\Bigg\}\Bigg] \, .
\end{eqnarray}

\begin{eqnarray}
\widetilde{\cal M}_{-2,2}&=&\sumintfff_{\{PQR\}}\frac{R^4}{P^2Q^2(P-Q)^4(Q-R)^2(R-P)^2}\nn
&=&-\frac{1}{(4\pi)^2}\left(\frac{T^2}{12}\right)^2
         \left(\frac{\Lambda}{4\pi T}\right)^{6\epsilon}\Bigg[\frac{1}{12\epsilon}\left(29+288\hat\mu^2-144
         \hat\mu^4\right) +\frac{89}{12}\nonumber\\ 
&&    +4\gamma_E + 2(43+24\gamma_E)\hat\mu^2   - 68\hat\mu^4+\frac{10}{3}\left(1+\frac{84}
        {5}\hat\mu^2\right)\frac{\zeta'(-1)}{\zeta(-1)} + \frac{8}{3}\frac{\zeta'(-3)}{\zeta(-3)}
        \nonumber\\
&&   + 24\left[10\aleph(3,z)+18i\hat\mu\aleph(2,z)+2(2-5\hat\mu^2)\aleph(1,z)
      +i\hat\mu\aleph(0,z)\right]\Bigg] \, .
\end{eqnarray}

\section{Three-dimensional integrals}
\label{app:threedints}

Dimensional regularization can be used to regularize both the ultraviolet divergences and infrared divergences in 3-dimensional integrals over momenta. The spatial dimension is generalized to  $d = 3-2\epsilon$ dimensions. Integrals are evaluated at a value of $d$ for which they converge and then analytically continued to $d=3$. We use the integration measure 
\begin{equation}
 \int\limits_p\;\equiv\;
  \left(\frac{e^{\gamma_E}\Lambda^2}{4\pi}\right)^\epsilon\;
\:\int {d^{3-2\epsilon}p \over (2 \pi)^{3-2\epsilon}}\;.
\label{int-def}
\end{equation}

\subsection{One-loop integrals}

The general one-loop integral is given by
\be\nonumber
I_n&\equiv&\int\limits_p{1\over(p^2+m^2)^n}\\
&=&{1\over8\pi}(e^{\gamma_E}\Lambda^2)^{\epsilon}
{\Gamma(n-\mbox{$3\over2$}+\epsilon)
\over\Gamma(\mbox{$1\over2$})
\Gamma(n)}m^{3-2n-2\epsilon}
\;.
\ee
Specifically, we need
\be\nonumber
I_0^{\prime}&\equiv&
\int_p\log(p^2+m^2)\\
&=&
-{m^3\over6\pi}\left({\Lambda\over2m}\right)^{2\epsilon}
\left[
1+{8\over3}
\epsilon
+{\cal O}\left(\epsilon^2\right)
\right]\;,\\ 
I_1&=&-{m\over4\pi}\left({\Lambda\over2m}\right)^{2\epsilon}
\left[
1+2\epsilon+{\cal O}\left(\epsilon^2\right)
\right]\;,\\
\label{i2}
I_2&=&{1\over8\pi m}\left({\Lambda\over2m}\right)^{2\epsilon}
\left[1+{\cal O}\left(\epsilon\right)
\right]
\;.
\ee

\subsection{Two-loop integrals}

We also need a few two-loop integrals on the form
\be
J_n&=&\int\limits_{pq}{1\over p^2+m^2}{1\over(q^2+m^2)^n}
{1\over({\bf p}+{\bf q})^2} \;.
\ee
Specifically, we need $J_1$ and $J_2$ which were calculated in refs.~\cite{3loopglue2}:
\be
J_1&=&
{1\over4(4\pi)^2}\left({\Lambda\over2m}\right)^{4\epsilon}
\left[
{1\over\epsilon}+2
+{\cal O}(\epsilon)
\right]\;,\\
J_2&=&
{1\over4(4\pi)^2m^2}\left({\Lambda\over2m}\right)^{4\epsilon}
\left[1+{\cal O}(\epsilon)
\right]\;.
\ee

\section{Properties of the $\aleph$ functions}
\label{app:aleph}

For some frequently occurring combinations of special functions we will apply the following abbreviations
\be
\zeta'(x,y) &\equiv& \partial_x \zeta(x,y) \, , \label{}\\
\aleph(n,z) &\equiv& \zeta'(-n,z)+\lb-1\rb^{n+1}\zeta'(-n,z^{*}) \, , \\
\aleph(z) &\equiv& \Psi(z)+\Psi(z^*) \, ,
\ee
where $n$ is assumed to be a non-negative integer and $z$ is a general complex number given here by $z=1/2-i\hmu$. Above $\zeta$ denotes the Riemann zeta function, and $\Psi$ is the digamma function
\be
\Psi(z)&\equiv&\frac{\Gamma '(z)}{\Gamma(z)} \, .
\ee
Below we list Taylor expansions of the function $\aleph(z)$ and $\aleph(n,z)$ for values of $n$ necessary for calculation of the susceptibilities presented in the main text.  For general application we evaluate the $\aleph$ functions exactly using \emph{Mathematica}.
\begin{eqnarray}
 \aleph(z)&=&-2\gamma_E-4\ln 2+14\zeta(3)\hmu^2-62\zeta(5)\hmu^4+254\zeta(7)\hmu^6+{\cal O}(\hmu^8) \, , \\
\nn
 \aleph(0,z)&=& 2\lb2\ln2+\gamma_E\rb i \hmu - \frac{14}{3}\zeta(3)i\hmu^3+\frac{62}{5}\zeta(5)i\hmu^5
+{\cal O}(\hmu^7) \, ,\\
\nn
 \aleph(1,z)&=&-\frac{1}{12}\lb\ln2-\Za\rb - \lb1-2\ln2-\gamma_E\rb\hmu^2-\frac{7}{6}\zeta(3)\hmu^4\nn
&&+\ \frac{31}{15}\zeta(5)\hmu^6
+{\cal O}(\hmu^8) \, ,  \\
\nn
\aleph(1,z+z') &=& -\frac{1}{6}\frac{\zeta'(-1)}{\zeta(-1)}-\lb1-\gamma_E\rb\lb\hmu+\hmu'\rb^2-
\frac{\zeta(3)}{6}\lb\hmu+\hmu'\rb^4\nn
&&+\ \frac{\zeta(5)}{15}\lb\hmu+\hmu'\rb^6+\mathcal{O}(\hmu^8,\hmu'^8)\, , \\
\aleph(2,z)&=&\frac{1}{12}\lb1+2\ln2-2\Za\rb i\hmu +\frac{1}{3}\lb3-2\gamma_E-4\ln2\rb i \hmu^3\nn
&&+\ \frac{7}{15}\zeta(3)i\hmu^5
+{\cal O}(\hmu^7) \, , \\
\aleph(2,z+z') &=& -\frac{1}{6}\Big(1-2\,\frac{\zeta'(-1)}{\zeta(-1)}\Big)i\lb\hmu+\hmu'\rb +
\frac{1}{3}\lb3-2\gamma_E\rb i \lb\hmu+\hmu'\rb^3\nn
&&+\ \frac{\zeta(3)}{15}i\lb\hmu+\hmu'\rb^5+\mathcal{O}(\hmu^7,\hmu'^7),
\end{eqnarray}
\begin{eqnarray}
\aleph(3,z)&=&\frac{1}{480}\lb\ln2-7\Zc\rb +\frac{1}{24} \lb5+6\ln2-6\Za\rb\hmu^2 
\nn
&&\hspace{1cm}
+\frac{1}{12}\lb11-6\gamma_E-12\ln2\rb\hmu^4+\frac{7}{30}\zeta(3)\hmu^6
+{\cal O}(\hmu^8) \, , \\
\nn
\aleph(3,z+z') &=& \frac{1}{60}\frac{\zeta'(-3)}{\zeta(-3)}-\frac{1}{12}\Big(5-6\,\frac{\zeta'(-1)}{\zeta(-1)}\Big)
\lb\hmu+\hmu'\rb^2\nn
&& +
\frac{1}{12}\lb11-6\gamma_E\rb\lb\hmu+\hmu'\rb^4+\frac{\zeta(3)}{30}\lb\hmu+\hmu'\rb^6+\mathcal{O}(\hmu^8,\hmu'^8) \, .
\end{eqnarray}


\end{document}